\def\hst{{\sl HST}}
\def\swift{{\sl Swift}}
\def\acs{{\sl ACS}}
\def\wfc3{{\sl WFC3}}
\def\uvot{{\sl UVOT}}

\def\galex{{\sl GALEX}}

\documentclass[12pt,preprint]{aastex}
\usepackage{appendix}
\usepackage{epsf}
\usepackage{color}
\usepackage{graphicx}
\usepackage{graphics}
\usepackage{epsfig}
\usepackage{longtable}
\usepackage{multirow}
\begin{document}
\title{The Panchromatic Hubble Andromeda Treasury. VII. The Steep Mid-Ultraviolet
  to Near-Infrared Extinction Curve in the Central 200 pc of the M31 Bulge}
\author{Hui Dong$^{1,2}$, Zhiyuan Li$^{3,4}$, Q. D. Wang$^2$, Tod R. 
  Lauer$^1$, Knut Olsen$^1$, Abhijit Saha$^1$, Julianne Dalcanton$^5$,
Karl Gordon$^{6,7}$, Morgan Fouesneau$^5$, Eric Bell$^8$, Luciana Bianchi$^9$}

\affil{$^1$ National Optical Astronomy Observatory,
Tucson, AZ, 85719}\affil{$^2$ Department of Astronomy, University of Massachusetts,
Amherst, MA, 01003}\affil{$^3$ School of Astronomy and Space Science, Nanjing University, Nanjing, 210093, China}\affil{$^4$
Department of Physics and Astronomy, University of California, Los
Angeles, CA, 90095}\affil{$^5$ Astronomy Department, University of Washington,
Seattle, WA, 98195}\affil{$^6$ Space Telescope Science Institute, 3700
San Martin Drive, Baltimore, MD, 21218}\affil{$^7$ Sterrenkundig
Observatorium, Universiteit Gent, Gent, Belgium}\affil{$^8$ Department of
Astronomy, University of Michigan, Ann Arbor, MI
48109}\affil{$^9$ Department of Physics and Astronomy, Johns Hopkins
University, Baltimore, MD 21218}\affil{E-mail: hdong@noao.edu}

\begin{abstract}
We measure the extinction curve in the central 200 pc of M31 at
mid-ultraviolet to near-infrared wavelengths (from 1928\AA\ to 
1.5$\mu$m), using \swift/\uvot\ and  
\hst\ \wfc3/\acs\ observations in thirteen bands. 
Taking advantage of the high angular resolution of the \hst\
\wfc3\ and \acs\ detectors, we develop a method to simultaneously
determine the relative extinction  
and the fraction of obscured starlight for five dusty complexes located
in the circumnuclear region. The extinction curves of these clumps
($R_V$=2.4-2.5) are steeper than the average Galactic one
($R_V$=3.1), but are similar to optical and near-infrared curves
recently measured toward the Galactic
Bulge ($R_V\sim2.5$). This similarity suggests that steep extinction
curves may be common in the inner bulge of galaxies. 
In the ultraviolet, the extinction curves of these clumps are
also unusual. We find that one dusty clump (size $<$2 pc) exhibits a
strong UV bump (extinction at 2175\AA ), more than three standard
deviation higher than that predicted by common models. 
Although the high stellar metallicity of the M31 bulge indicates
that there are sufficient carbon and silicon to produce large dust
grains, the grains may have been destroyed by supernova explosions or past activity
of the central super-massive black hole, resulting in the observed steepened extinction curve. 
\end{abstract}

\section{Introduction}\label{s:intro}
Dust grains are pervasive in the Universe, absorbing, scattering and
re-radiating light, affecting all wavelengths. Accounting for the effects of dust is 
one of the fundamental steps when inferring intrinsic
properties of astrophysical objects. The degree
of the effects depends not only on the total
 column density of dust grains, but also on their size and 
composition. Dust grains of various sizes affect different parts of the
electromagnetic spectra. Small grains mainly absorb at
 shorter wavelengths, such as the 
ultraviolet (UV), while large grains dominate attenuation in the
infrared (IR). In particular, carbonaceous grains 
are suggested to cause strong extinction near
2175\AA\ \citep{dra03}. The overall wavelength dependence of dust
extinction is called the `extinction law' (or extinction curve), which is
conventionally expressed to be the ratio between the
absolute extinction, $A_{\lambda}$, at some wavelength, 
$\lambda$ and the absolute extinction in the $V$ band, $A_V$, as a function of the reciprocal of
the wavelength. The extinction curve is governed by the mix of dust grains,
which can potentially be affected by the local environments. 
Strong shocks and UV photons
could destroy large grains and thus change
the shape of the extinction curve~\citep{jon04}. 

Extinction curves have been extensively studied in the
Milky Way (MW,~\citealt{fit04} and reference therein) and in the Magellanic Clouds 
(MCs, Large Magellanic Cloud: LMC and
Small Magellanic Cloud: SMC,~\citealt{gor03}). 
Thanks to the International Ultraviolet
Explorer (IUE), many high-quality low-resolution UV spectra of the
stars in the MW and the MCs have made previous work on
extinction curves possible. These studies have revealed significant
environmentally-dependent effects on the extinction curves, reflected
in varying UV slopes and strengths of the 2175\AA\
bump.~\citet{car89} find that most
 extinction curves in the MW could 
be expressed with a function that depends on a 
single parameter, $R_V$=$A_V$/($A_B$-$A_V$) ($A_B$ is the absolute
extinction in the $B$ band), which
roughly traces the dust grain size. Cardelli's
extinction curve steepens (i.e. the relative extinction in the short
wavelength becomes large) with decreasing $R_V$, although
deviations are found toward several directions~\citep{mat92}. One of the
most significant features of the MW extinction curve is the strong
2175\AA\ bump, the width of which is sensitive
to the local environment (from 0.63 to 1.47 $\mu m^{-1}$;
~\citealt{val04}). 
In contrast to the well-behaved extinction curves in the MW, 
the extinction curves in the MCs, especially the SMC, are much steeper
in the UV bands and exhibit a significantly weaker 2175\AA\ bump~\citep{gor98,mis99}.~\citet{gor03} fit the extinction curves
in the MCs with the generalized model provided by~\citet{fit90}
(similar to that of~\citealt{car89}), and
claim that the variation in dust properties in the MW and MCs is 
caused by environmental effects. 

The Andromeda galaxy~\citep[M31, at a distance
of $\sim$780 kpc;][]{mcc05} provides us with an ideal testbed 
to study the extinction curves in regions with different
metallicity and star-forming activity. The extinction curve in the M31 disk 
is similar to the `average' Galactic one ($R_V$=3.1), albeit
 with a possibly weaker 2175\AA\ bump~\citep{bia96}. Using ground-based optical images in
BVRI bands, \citet{mel00} find that the extinction curve of a dusty
complex 1.3\arcmin\ on the sky ($\sim$300 pc in projection) northwest of the M31 nucleus is 
much steeper ($R_V\sim2.1$). 

In this work, we study the extinction curve in the central 200 pc of 
the circumnuclear region (CNR) of M31. As the second closest galactic
nucleus, the CNR of M31
offers a unique laboratory~\citep[][and references therein]{li09} for studying the interaction and co-evolution 
between the super-massive black holes (SMBHs) and their host galaxies.
By virtue of proximity, we can achieve an unparalleled linear resolution
in M31 for a detailed study on 
various astrophysical activities in an extreme 
galactic nuclear environment. Like our Galaxy, M31 harbors a
radiatively quiescent
SMBH, named M31*~\citep{dre88,kor88,cra92,gar10,li11}. 
On the other hand, unlike the active
star formation in the Galactic Center, the nuclear
bulge of M31 does not host any young massive stars (less than 10
Myr old,~\citealt{bro98,ros12}) and contains only a small amount of molecular 
gas~\citep{mel00,mel11,mel13}. The stellar population in the M31 bulge is found to be
highly homogeneous, dominated by old stars 
($\sim$8 Gyr,~\citealt{ols06,sag10}).
In the central 2\arcmin\ ($\sim$450 pc) , the two-dimensional
surface brightness distribution of the bulge agrees well with a S\'ersic
Model~\citep{pen02,li13}. The metallicity in the M31
bulge seems to be super-solar~\citep{sag10} and much higher than that of the MCs. 
The steep extinction curve claimed by~\citet{mel00} may be due
to the nuclear environment of the galaxy,
with its high metallicity, 
as well as the potential impact of the SMBH (i.e. due to
ongoing mechanical feedback and/or previous outbursts) and strong
 interstellar shocks, all of which could affect the size 
and compositions of the dust
grains. 

Because of the relatively low line-of-sight 
extinction, the CNR of M31 is the nearest well-defined
galaxy nucleus that can be mapped 
from the UV to the near-IR (NIR) bands. To our knowledge, there has not
yet been a study of the extinction curve covering the UV-optical-NIR
wavelength range in the central $\sim$ 500 pc of a normal galaxy. The 
understanding of the extinction curve over this entire range is
essential to studies of distant galactic nuclei, especially
 for those with similar properties.

In this paper, we empirically derive the relative extinctions at 13
bands from the mid-UV (MUV) to NIR and then determine the 
extinction curves for
representative regions in the CNR of M31. We utilize 
data from the \hst\ Wide Field Camera 3 ({\sl WFC3}) and Advanced Camera for
Surveys (\acs ) of multiple programs~\citep{dal12,li13}. 
The cores of dusty clumps can be
resolved in our data, thanks to the superb angular resolution of \hst\ ($<$0.15\arcsec,
i.e. $\sim$0.55 pc), while the high sensitivity of the \hst\ \wfc3\
and \acs\ cameras ensures 
high signal-to-noise (S/N) ratios. We
also utilize \swift/\uvot\ observations with three MUV filters, the middle of which covers the 2175\AA\ bump. Therefore, the \swift/\uvot\ 
filters can be used not only to examine the slope of the extinction
curve in the MUV, but also to probe the strength of the 2175\AA\
bump. In a companion
work,~\citet{li13} studied the fine spatial structures of the extinction 
features in the CNR of M31. 


We present the \swift\ and \hst\ 
observations, and the data reduction in
\S\ref{s:observation}. We describe our 
method to derive the line-of-sight locations and the extinction curves in
\S\ref{s:method}, apply it to the dusty clumps in M31's CNR in
\S\ref{s:analysis} and present the results in \S\ref{s:result}. We
discuss the implications of our results in
\S\ref{s:discussion} and conclude the paper in \S\ref{s:summary}. 

\section{Data}\label{s:observation}
\subsection{\hst\ \wfc3\ and \acs\ observations}\label{ss:hst_obs}
We utilize images taken by \hst\ \wfc3\ and \acs\ in ten bands (see
Table.~\ref{t:obs}); six bands from the Panchromatic Hubble Andromeda
Treasury Survey (PHAT;~\citealt{dal12}, GO-12055), three bands from Program
GO-12174~\citep{li13}, and one band from each of 
Programs GO-10006, GO-10760 and GO-11833 (PI: Michael Garcia). 
The central wavelength of the \wfc3\ F547M filter (5447 \AA )
is close to that of the traditional Johnson $V$ band (5400 \AA ). Therefore, we
consider F547M as an analog of the $V$ band in this work. The detailed description of these programs and data
reduction steps are
given in~\citet{don13} and~\citet{li13}, but we describe the key steps
here. We correct for bad pixels, dark current and flat-fielding for
individual dithered exposures with `calwfc3' and `calacs' in
`PyRAF'. We then merge exposures of each pointing
position using `Multidrizzle'\footnote{`PyRAF' and `Multidrizzle' are the product of 
the Space Telescope Science Institute, which is operated by AURA for
NASA.}. We first correct for the relative astrometry among the F475W images at different
pointing positions, using the $\chi^2$ minimization, as described 
in~\citet{don11}. The images of the other
nine filters are aligned to the coordinate system of the F475W
band. We calculate relative bias offsets of the position images in individual
bands, using the same $\chi^2$ minimization method. 
Because the CNR of M31 has a high surface brightness in all
 ten bands, we do not subtract a sky background. Using the aligned,
 bias corrected images, we then construct a
 mosaic image for each band. We match the resolution of each of the
 other nine filters to the poorest resolution of F160W band (FHWM,
$\sim$0.15\arcsec), by utilizing point spread functions (PSFs) of
 the ten bands produced by
 `Tinytim'\footnote{http://www.stsci.edu/hst/observatory/focus/TinyTim}
 and `PSFMATCH' in `PyRAF' to produce the appropriate kernels. 
All the final images are rebinned to 0.13\arcsec /pixel ($\sim$ 0.5 pc
/pixel), the 
pixel size of the WFC3/IR detector. 

We use a box filter to empirically determine local background and
 intensity uncertainty maps, as well as to remove distinct stellar sources in our
 analysis  (\S\ref{s:analysis}). The size of the box is
5$\times$5 pixels ($\sim$0.65\arcsec$\times$0.65\arcsec ,
i.e. 4.3$\times$4.3 F160W FWHM), which is large enough to remove
sources and small enough to trace local
background fluctuations. The median and 68\% percentage uncertainty
within the box in the mosaic of each filter ($n$) are used to represent
the local background ($B_n$) and its associated standard deviation
 ($\sigma_n$) of the central pixel of this
box. The median ratios of the
uncertainty to the intensity within the
0.65\arcsec$\times$0.65\arcsec boxes are listed in
Table~\ref{t:obs} for the ten \hst\ bands. The F275W, F110W and F160W
bands have the lowest signal-to-noise ratio (SNR), because of either the
short exposure time
(F275W) or the stochastic uncertainty due to the presence of unresolved
red giant branch (RGB) and asymptotic giant branch (AGB) stars, which are bright in the near-IR bands (F110W and
F160W). We
identify the pixels with intensity larger than $B_n$+3$\sigma_n$ in
at least one of the ten \hst\ bands, as `source' pixels. We remove each of these pixels 
as well as their immediate neighbors (i.e., a 3x3 box, $\sim$
2.5$\times$2.5 F160W FWHM)  to avoid the contamination from the
wing of bright sources. 


\subsection{\swift /\uvot\ Observations}\label{ss:swift_obs}
\swift /\uvot\ is a 30 cm UV and Optical telescope on board the \swift\
spacecraft~\citep{rom05}. 
In this work, we use the three MUV filters of \swift
/\uvot : UVW2, UVM2, UVW1 (see Table~\ref{t:obs}). 
The UVM2 filter encompasses the 2175\AA\
bump, while the other two
filters (UVW2 and UVW1) 
cover the blue and red sides of the bump, respectively. These three
filters have a strong effect on constraining extinction curves, because various extinction
curves have the largest differences in the slopes of the extinction
curves in the UV band and in the strength of the 2175\AA\ bump. UVW1 
is close to \wfc3\ F275W. Because UVW2 and
UVW1 have extended red tails (the `Red Leak' problem) and the M31
bulge is bright in the optical and the NIR, 
the effective wavelengths and thus the relative extinction
($A_{\lambda}$/$A_V$) of the two bands are sensitive to 
the age and metallicity of
 the background stellar populations (see Appendix
A). In \S\ref{s:result}, we describe a method to compare the
relative extinction derived from our observation with the ones predicted by different
extinction curves. The angular resolution of \swift /\uvot\ in the MUV 
($\sim$2.5\arcsec) is poorer than that of
\hst\ observations ($<$0.15\arcsec), but is still better than
\galex\ (4\arcsec - 5\arcsec)~\citep{mor07}. To reduce the amount of telemetry from the spacecraft, all of
the observations in image mode have already been binned by a factor of 2, with a
pixel scale of 1\arcsec\ pixel$^{-1}$ ($\sim$3.8 pc pixel$^{-1}$). 
After excluding several observations
with exposure time shorter than 20 seconds, the total exposures times
are 106, 46 and 153 kilo-seconds for the UVW2, UVM2 and UVW1
filters, respectively. 

We process the data using the steps described in the UVOT Software Guide. First, we reproduce the level I products with the
 UVOT FTOOLS (HEAsoft 6.12). Second, we rectify the `coincidence
loss' problem (a problem in which the UVOT camera counts 
only one photon, even if there are more than 
one photon arriving at the same CCD pixel in one readout frame) by using Fig. 6 of~\citet{bre10}. 
We remove pixels with intensity $>$0.4 counts $s^{-1}$
pixel$^{-1}$, for which the correction for coincidence
loss is large and uncertain. Fortunately, we find that only
the central $<$ 5\arcsec\ (19 pc) of M31 and the cores of several bright
foreground stars are removed, even in the UVW1 band, which has
the biggest chance of suffering from the coincidence loss
problem. Third, we manually remove the `smoke
rings' found in the UVW2 and UVW1 observations, using 
an annulus with 30\arcsec\ and 140\arcsec\ for 
the inner and outer radii (see also Fig. 20 of~\citealt{bre10}).
 Fourth, we derive the relative astrometry between individual
exposures of each filter, using $\chi^2$ minimization as in~\citet{don11}
to correct for the relative astrometry between individual
exposures of each filter before producing final mosaics. 
The bright UV stars in the \swift\ observations are used to
align the final mosaics to the coordinate system of the \hst\
observations. Fifth, we construct the PSF for the three filters, using 
relatively bright UV stars (but
without the `coincidence loss' problem). 

\section{Method}\label{s:method}
In this section, we describe our method to constrain the
extinction curves of individual dusty clumps identified in the CNR of M31. 
 We first review two conventional methods and their limitations 
if they were applied to  the M31 bulge, and a third method adopted in
our companion work~\citep{li13}. The widely used 
`standard pair' method~\citep{mas83} compares the spectra of pairs of stars of a similar
 spectral type, but one with high absolute extinction and the other
 not. Early-type
stars (usually O or B) are generally used, because they are bright in 
the UV.~\citet{bia96}, for example, apply this method to young massive
stars in some OB
associations of the M31
disk. The method cannot be used here, because such stars are
absent in the CNR of
M31~\citep[e.g.,][]{bro98,ros12}. 


An alternative method is to use integrated light to derive the
extinction curve. \citet{elm80} is among the first to employ such 
a method in several spiral
galaxies. \citet{wal88} and \citet{mel00} apply a similar method to 
study dust extinction in the disk and the bulge of M31, respectively. In particular,~\citet{mel00} focus on the dust complex
D395A/393/384, located at $\sim$1.3\arcmin\ northwest of the M31* and
 derive  the optical extinction curve, $\langle A \rangle$/($\langle A_B\rangle$-$\langle A_V\rangle$), from the mean extinctions
($\langle A\rangle$) in four bands (BVRI). However, this work is forced to 
assume the fraction, $f$, of obscured
starlight, due to the lack of the
information about the line-of-sight locations of this dusty
clump.
Recently, \citet{li13} use the same \hst\ dataset here to directly 
constrain the pixel-by-pixel values of $f$ for various dusty
clumps in the M31's CNR, assuming that they 
are all located in a thin plane embedded in a triaxial ellipsoid bulge~\citep{sta77}. 

Here, we employ an alternative approach, relaxing the geometric
assumptions made by~\citet{li13}. Because the filling factor of 
dusty clumps is low (see~\citealt{don13,li13}), it is
reasonable to assume that they are relatively isolated individual
features, with each cloud occupying a single depth inside the bulge and that
their sizes (and hence thicknesses) are small compared to the bulge
depth. Therefore, for a single dusty
clump, the fraction of starlight behind the
dusty clump can be treated as a constant and in our approach, we
consider a single $f$ value per each dusty clump. 

For one pixel in the mosaic 
of the $n^{th}$ filter, we calculate the ratio $\Re_n$ between the observed
intensity $I_n$, the intrinsic intensity $S_n$, given by 
the fraction of obscured starlight $f_n$ and the
absolute extinction\footnote{Due to the broadness of the
  filters, their effective wavelengths are sensitive to the background
stellar light (see Appendix A). We use $A_n$, instead of
$A_{\lambda}$ to represent the absolute extinction within the $n^{th}$ filter.} $A_n$ in
Eqn.~\ref{e:ratio}. 
\begin{equation}\label{e:ratio}
\Re_n=\frac{I_n}{S_n}=(1-f_n)+f_n\times10^{-0.4\times A_n}
\end{equation} 
The terms $(1-f_n)$ and $f_n\times10^{-0.4\times A_n}$ are the fractions of
unobscured and obscured starlight, respectively. $I_n$ is from the observations in
\S\ref{s:observation} and we derive $S_n$, using the method detailed in
\S\ref{ss:intri}. This equation neglects the starlight scattered
into the line of sight, which we show is a safe assumption in Appendix C. 
In Fig.~\ref{f:f_a_rela}, we depict the non-linear relationship 
between $f_n$ and $A_n$ for different $\Re_n$. We see that $f_n$ is
anti-correlated with $A_n$. At small $f_n$, $A_n$ is sensitive to $f_n$, but changes
slowly when $f_n>0.8$. 
For a fixed $\Re_n$ value, simply assuming $f_n$=1 for a dust clump in 
very extinguished regions, we could potentially bias
ourselves towards underestimating the actual extinction, $A_n$.

We can convert Eqn.~\ref{e:ratio} into a canonical extinction curve by
normalizing the extinction in band $n$ to the extinction in our
nominal $V$-band proxy, F547M, assuming that the extinction
curve is constant within each dusty clump. If we define the ratio of
extinctions as $\frac{A_n}{A_{F547M}}\equiv\Gamma(n,F547M)$
[or $\Gamma$(n) for simplicity, unless otherwise noted] and assume
that $f$ is the same for all filters ($f_n=f_{F547M}=f$), we can
eliminate
 the absolute extinction $A_{F547M}$ to obtain the
following equation for each individual pixel ($k$):
\begin{equation}\label{e:f_a}
\Re_n-1+f=f^{1-\Gamma(n)}\times(\Re_{F547M}-1+f)^{\Gamma(n)}
\end{equation} 
We can then estimate, the fraction of obscured starlight, $f$, and
the relative extinction, $\Gamma(n)$, for each dusty clump, 
by minimizing the following: 
\begin{equation}\label{e:indi}
\chi^2=\sum_{k}\sum_{n}\frac{[\Re_n-1+f-f^{1-\Gamma(n)}\times(\Re_{F547M}-1+f)^{\Gamma(n)}]^2}{(\frac{\sigma_n}{I_n})^2+(\frac{\sigma_{F547M}}{I_{F547M}})^2+(\frac{\delta
  S_n}{S_n})^2+(\frac{\delta S_{F547M}}{S_{F547M}})^2}
\end{equation}
where $\sum_{k}$ and $\sum_{n}$ are the sum over all the pixels
and over the included bands (see \S\ref{s:analysis}) for a given dusty clump. The variables
$\sigma_n$ and $\sigma_{F547M}$ are the 
photometric uncertainties of
$I_n$ and $I_{F547M}$, respectively, as
determined in \S\ref{ss:hst_obs}. The variables $\delta  S_n$ and $\delta
S_{F547M}$ are the uncertainties of $S_n$ and
$S_{F547M}$, determined in \S\ref{ss:intri}. 

\section{Analysis}\label{s:analysis}
We apply the above method to the \swift/\uvot\ and \hst\ data.
We are primarily interested in the dusty clumps within the central
10\arcsec\ (38 pc) to 60\arcsec\ (227 pc), a region that has drawn relatively little 
attention in previous studies. For each clump, we assume
that the obscured 
fraction $f$ is a constant among the different
bands~\citep{li13}, the included
satisfying the requirement of Eqn.~\ref{e:f_a}.

In Fig.~\ref{f:ratio}, we show the \hst/\wfc3\ F336W
 intensity map (Fig.~\ref{f:ratio}a) and the intensity ratio map between F160W
 and F336W (Fig.~\ref{f:ratio}c) in the central 2\arcmin$\times$2\arcmin\ of
the M31 bulge. This ratio is sensitive to 
extinction, as well as age and metallicity of local stellar
populations, in the sense that large extinction, old age, or high
metallicity could enhance this ratio. The majority of the field of view
seems to be 
free of patchy extinction. We use cyan boxes to mark five dark
and fuzzy structures in the intensity ratio map; these structures have low
F336W intensity and should indicate sites of genuine high
extinction. The locations and sizes of these five cyan boxes are given
in Table.~\ref{t:clumps}. Due to their low surface density
($<10^{21}$ {\rm H cm}$^{-2}$,~\citealt{li13}), these five regions do not have
available CO detections yet. In the extinction map presented by~\citet{li13}, these
five regions include many high extinction regions that appear isolated 
from each other. The high intensity ratio in the central 10\arcsec\
is more likely caused by the nucleus having a stellar
population different from that of the rest of the M31 bulge
(see~\citealt{don13}), rather than significant extinction variations
given that previous studies do not find molecular clouds in the same
central 5\arcsec\ (19 pc) region of M31. 


We first describe how to construct the
intrinsic surface brightness distribution of the M31
bulge in \S\ref{ss:intri} and in \S\ref{ss:mole}, we define our
selection criteria on individual pixels to define dusty clumps. We
then minimize the $\chi^2$ (Eqn.~\ref{e:indi}) in two steps, first
using only \hst\ images (\S\ref{ss:f_a}). We then extend in
\S\ref{ss:f_a_uvot} the minimization to including \swift\ images after
correcting for the unresolved sources and differential extinction
within the larger pixels of \swift\ based on \hst -only derived
values. Finally, we address several caveats of our analysis in \S\ref{ss:caveats}.


\subsection{Intrinsic Light Distribution in the M31 Bulge}\label{ss:intri}
We fit the surface brightness distribution, using the surface
photometry algorithm described in~\citet{lau86} to determine $S_n$ for each
filter. This algorithm is similar to the `{\tt ellipse}' task in `IRAF'~\citep{jed87}. 
It first divides an image into a set of
concentric annuli centered on M31*. The width of these annuli
increases by 0.1 in a log scale. The surface brightness distribution
of each annulus is fit
with an ellipse. The normalization, ellipticity, and position angle of
each annulus is solved simultaneously, using a 
non-linear $\chi^2$ minimization method. Similarily to sigma-clipping,
we iteratively flag and then mask pixels of dusty clumps, detector artifacts and discrete
sources to refine our fit of the intrisic light profile. After the
parameters of ellipses for individual annuli are derived, we 
recover the intrinsic surface brightness distribution, $S_n$, 
in each filter image. Because of the large number of pixels and high
SNR data, the statistical uncertainty of $S_n$, within a
given annulus, $\delta S_n$, is typically of 
0.2\%, much smaller than the photometric uncertainty of a single pixel
(i.e. $\sigma_n$). In Fig.~\ref{f:IvsS}, we depict the distribution of
$I_n$/$S_n$ in the central 2\arcmin$\times$2\arcmin\
($\sim$450 pc$\times$450 pc) of the M31 bulge in the F275W, F547M and
F814W bands. The respective dispersions of the distributions of
$I_n$/$S_n$ (0.086, 0.035 and 0.058) are similar to
the photometric uncertainties in the three bands ($\sigma_n$, see
Table.~\ref{t:obs}). 

\subsection{Identifying Pixels Associated with Dusty Clumps}\label{ss:mole}
When fitting the extinction curve, we select only those pixels with strong extinction,
which we refer to as `dusty' pixels. For each band, the histogram of
the pixel values in the $\Re_n$ map within the central 
2\arcmin$\times$2\arcmin\ ($\sim$450 pc$\times$450 pc) is well
fit with a Gaussian distribution. Because 
the filling factor of the dusty clumps
is small in the M31 bulge, the centroid of the Gaussian distribution
is around one 
and the standard deviation is dominated by the photometric
uncertainty. The extra pixels at the wings of the Gaussian curve are
due to bright sources (large $\Re_n$) or high extinction (small
$\Re_n$), which cannot be reproduced by the method in
\S\ref{ss:intri}. The `dusty' pixels are chosen to have 
$I_n/S_n$ two standard deviations below one  
in all the UV and optical bands (i.e., F275W, F336W,
F390M, F435W, F475W, F547M). If $f$=1, using Eqn.~\ref{e:ratio}, 
this threshold then corresponds to $A_{F275W}>$ 0.2 or 
$A_{F547M}>$ 0.04. With the threshold, fewer than 
1.5\% of the pixels are expected to be randomly below this limit and
thus the pixels passing this cut almost certainly suffer strong 
extinction. In contrast, in the images at longer wavelength, the ratio
is mostly determined by the photometric uncertainty and only a 
few pixels with very high extinction could have ratios significantly below
one. The same `dusty' pixels are used for
all the filters in the following analysis. 


\subsection{Constraining $f$ with the \hst\
  Observations}\label{ss:f_a}
We use Eqn.~\ref{e:indi} to derive the fraction of obscured
starlight, $f$, and the extinction relative to the $V$ band, $\Gamma(n)$, for the five dusty
clumps as defined in Fig.~\ref{f:ratio}c. 
Using the `MPFIT' package~\citep[Gradient descent,][]{mar09}, we simultaneously fit 
the F336W, F390M, F435W, F475W bands, since the intensities in these
bands are most sensitive to the extinction. The F275W band is excluded
because of the large statistic photometric uncertainty (see
Table~\ref{t:obs}). Because $\Re_n$ is typically close to 1 at
  near-IR wavelengths, the fraction of obscured starlight,
$f$, is very insensitive to the absolute extinction. Including the
longest wavelength bands in the fitting will
introduce large uncertainty into $f$ and thus into the relative
extinction. A Monte Carlo method is then used to evaluate the uncertainties in $f$
and $\Gamma(n)$. For each pixel, we randomly add a value which
follows a normal distribution with a mean of zero and a standard
deviation equal to $\frac{\sigma_n}{S_n}$ and
$\frac{\sigma_{F547M}}{S_{F547M}}$ into $\Re_n$ and
$\Re_{F547M}$, respectively. We rerun `MPFIT' 1000 times to obtain 
the 68\% percentage uncertainty of the $f$ and $\Gamma(n)$, as
listed in Table~\ref{t:f_a_uvot}.


Fig.~\ref{f:demonstrate} illustrates two examples of our fitting
results for Clump A and Clump D. The top panels show $\Re_{F336W}$
versus $\Re_{F547M}$ for all the `dusty' pixels in each cloud. The
bottom panels show the $\chi^2$ distribution as the function of $f$ and
$\Gamma(F336W)$. The best-fit parameters show that the two clouds
have similar $\Gamma(F336W)$, but different $f$. To show the
sensitivity of the data to different values of $f$ and
$\Gamma(F336W)$, in the top panels of Fig~\ref{f:demonstrate}, we also 
compare the observed data points with the curves predicted by various $f$ and
$\Gamma(F336W)$. For those `dusty' pixels with relatively low
absolute extinction  (such as $\Re_{F547M}>0.85$), the relationship
between $\Re_{F336W}$ and $\Re_{F547M}$ is not
sensitive to $f$, but constrains the $\Gamma(F336W)$ well; Instead, the
constraints on $f$ come purely from the most attenuated pixels. 

The derived $f$ could be used as an indicator of the line-of-sight
locations of the dusty clumps in the M31 bulge. 
Our results show that the two dusty clumps (Clump A and D, $f$ $>$ 0.5)
 are in front of the M31*, 
whereas the other two (Clumps B and E, $f$ $<$ 0.5) 
are behind, as summarized in Table~\ref{t:f_a_uvot}. Clump C ($f$$\sim$0.5) is at the middle.
 The uncertainties of $f$ for Clump C and Clump D 
are relatively large, because the former has the smallest number of
`dusty' pixels, whereas the latter does not have high enough
extinction (see Fig.~\ref{f:f_a_rela}). 


Using the `best-fit' values of $f$, we derive
$A_{F547M}$ from Eqn.~\ref{e:ratio}. The median $A_{F547M}$ values 
and their 68\% quantile for the five dusty clumps are 
0.25$\pm$0.10, 0.50$\pm$0.18, 0.31$\pm$0.08, 0.18$\pm$0.08 and
0.56$\pm$0.20, respectively. The histograms of $A_{F547M}$ for these five dusty clumps are presented in
Fig.~\ref{f:nh}. The $A_{F547M}$
distributions of Clumps B and E, having more high extinction pixels,
appear flatter than those of Clumps A, D and E. 


We fix the value of $f$ in
Eqn.~\ref{e:indi} and use `MPFIT' to derive the relative extinction 
$\Gamma(n)$ for the other \hst\ filters (F275W, F665N, F814W,
F110W, F160W), 
using the same Monte Carlo method to derive the
uncertainty of $\Gamma(n)$. We computed the 
effect of uncertainty in $f$ on $\Gamma(n)$ by deriving 
the values of $\Gamma(n)$ corresponding to the best-fit
$f$$\pm$$\delta$$f$ ($\delta$$f$ is the uncertainty in $f$). 
We then add half of the difference between the derived 
values of $\Gamma(n)$ in quadrature to the uncertainty 
derived from the Month Carlo method with $f$ fixed at
 its best-fit value to give the total variance in $\Gamma(n)$. The results are listed in
Table.~\ref{t:f_a_uvot}. 


\subsection{The Correction for \swift/\uvot\ Data}\label{ss:f_a_uvot}


We next use the \swift /\uvot\ images at the three MUV bands 
to constrain the extinction curve at shorter wavelengths. 
We first match the \hst/\acs\ PSF to that of \swift/\uvot . 
Since the PSF of the \swift/\uvot\ is much larger than 
that of the \hst\ (see \S\ref{s:observation}), we simply convolve the
\hst\ mosaics with the PSF of \swift/\uvot\ derived in
\S\ref{ss:swift_obs}. We still use the same bands in \S\ref{ss:mole}  to 
select out the `dusty' pixels. 
The intrinsic intensity distributions at these
three UV filters are also derived using the method of~\citet{lau86}. 


The contribution from unresolved sources in the three \swift/\uvot\
filters could potentially bias the extinction curve in the MUV
bands. Due to the relatively poor angular resolution and low
signal-to-noise ratio, 
 no `source' pixel is found in \swift/\uvot\ images of the five dusty
clumps. Moreover, because the M31 bulge is relatively dim
in the MUV bands, single UV bright stars can
dominate the integrated intensities of individual pixels. As a result,
we would
overestimate $\Re_n$ in the MUV bands and underestimate the
extinction, leading to flattened extinction curves. 
To minimize this effect, we first use the \hst\ F275W band, which
covers a comparable wavelength, 
 to estimate the source contribution in each pixel of
the \swift/\uvot\ images. We downgrade the resolution of the {\sl
  observed} F275W band images with and without the `source' pixels
 defined in \S\ref{ss:hst_obs}  by convolving them with the 
\swift/\uvot\ PSF and rebinning into the pixel size of the
\swift/\uvot\ observations (1\arcsec\ {\rm pixel}$^{-1}$). 
The ratio map between these two images represents the
fraction of `bright source-free' emission in individual \swift/\uvot\
pixels. The mean contribution of the point sources in the fields of the five
dusty clumps is between 27\% to 42\%. We then multiply the
corresponding ratio map to the observed \swift/\uvot\ 
UV images to remove the `source' contamination. To make sure that we
compare the observed and intrinsic intensities at the same sky area, 
we also need to remove the intensity 
contribution of the `source' pixels from individual pixels in the
intrinsic \swift/\uvot\ UV images. We produce a similar ratio map
for the intrinsic F275W image, $S_n$, (see
\S\ref{ss:intri}), which is then multiplied to the intrinsic intensity distributions of the
\swift/\uvot\ UV images. The resultant observed and intrinsic `bright source-free' 
intensity images are used to derive the $\Re_n$. After these corrections for point
source contamination, the $\Re_n$ of the \uvot\ bands decreases, the
relative extinction compared to the $V$-band 
increases by a factor of 1.2-1.8 and the extinction curve becomes steeper. 

Differential extinction within one \swift/\uvot\ pixel could also flatten
the extinction curve at the short wavelengths~\citep[`flattening bias'; see
also][]{cal00}, due to unresolved structures in the dust below the
angular scale of the pixel. From the \hst\ images, we know that the extinction can
change sharply on arcsecond scales (see Fig.~\ref{f:ratio} 
and~\citealt{don13,li13}). The low resolution of \swift/\uvot\ 
tends to lose this spatial information. As a result, the intensity in a
\swift/\uvot\ pixel may be dominated
by regions with the least 
extinction. If we just simply derived the extinction from the
$\Re_n$ (i.e. Eqn.~\ref{e:ratio}), we would average the
extinction within one \swift/\uvot\ pixel weighted by 
their observed intensity, and thus we would 
underestimate the local extinction. This effect would 
lead to the flattening of the inferred extinction curve, 
especially at short wavelengths. We correct for this bias in the three UV
bands, using the method
of~\citet{fri11}. The basic idea is that we can use the high resolution
extinction map of the
adjacent \hst\ filters, such as F336W, to mimic those of the three MUV
filters. The method is detailed
in Appendix B. The results, i.e. $\Gamma(n,F336W)$ for UVW2,
UVM2 and UVW1 are listed in
Table~\ref{t:uv_rela}. After this correction, $\Gamma(n,F336W)$ 
increases from $\sim$1.02 (Clump D) to $\sim$1.3 (Clump B). 
$\Gamma(n)$ is derived by
multiplying $\Gamma(n,F336W)$ to $\Gamma(F336W, V)$ 
and is listed in Table~\ref{t:f_a_uvot}. The errors of
$\Gamma(n,F336W)$ and $\Gamma(F336W, V)$ are 
propagated into the uncertainty of $\Gamma(n)$. 

\subsection{Caveats}\label{ss:caveats}
We describe four caveats of the analysis procedure described above:
the foreground
Galactic extinction, the filter dependence of $f$, the
scattered light and reddening substructures on scales below the \hst\
resolution. All the errors introduced by these caveats are summarized
in Table.~\ref{t:error}. 

\subsubsection{Foreground Galactic Extinction}\label{ss:fore}
What effect may the foreground Galactic extinction have on our measurements? In \S\ref{ss:intri}, 
we estimated the intrinsic intensity, $S_n$, in the field of the dusty clumps from the apparently  
extinction-free regions in the CNR of M31, which are, however, subject to  
Galactic extinction. Provided that
the Galactic extinction does not vary significantly
 within the field-of-view, this effect cancels in
$\Re_n$=$\frac{I_n}{S_n}$. The choice of high extinction pixels in
\S\ref{ss:mole} also
reduces the effect introduced by potential variation in the Galactic
extinction between $I_n$ and $S_n$ regions. The median $A_{F547M}$ of the `dusty' pixels in the clumps 
are larger than 0.18 mag. By constrast, the Galactic absorption toward the
M31 nucleus is 0.17~\citep{sch11}. Therefore, even a 10\% 
fluctuation of the foreground Galactic extinction within 
the central region of the M31 bulge would cause only 
$<$10\% error in $A_{F547M}$. 

\subsubsection{Dependence of the Fraction of 
Obscured Starlight on Wavelength}\label{ss:d_f}
The fraction of obscured starlight $f$ could vary among the
filters, because of potential stellar color variation across 
the M31 bulge. Such variation among different filters
($\arrowvert$$\frac{f_n-\langle f_n\rangle}{\langle
  f_n\rangle}$$\arrowvert$, where $\langle f_n\rangle$ is the mean
of $f_n$ for each of the dusty clumps) appears to be small. The maximum 
difference among the ten \hst\ filters is 
2.7\%, 7.1\%, 1.8\%, 4.7\% and 5.7\% for Clump A, B, C, D
and E, respectively. The relatively large variation seen in Clump B is
due largely to the F275W
band. Excluding it, the variation of $f_n$ decreases to 4.9\%. 
Therefore, the F275W band provides a good test of
the effect of $f$ on the relative extinction $\Gamma(F275W)$. We find
that if we were to increase  
$f$(F275W) by 7\%, $\Gamma(F275W)$ would decrease to
2.79 from 2.99, or a change of only 6.7\%, which is well within the uncertainty
range of $f$ of the clump.  

\subsubsection{Scattered Light}\label{ss:scattering}
The observed intensity is the sum of the
absorbed (background), unabsorbed (foreground) and scattered stellar light. The scattered
intensity depends on dust optical depth and scattering properties
(albedo and scattering phase function). 
In Appendix C, we run a simulation. We put a dusty clump with
different column densities and locations into the M31 bulge to derive the
fraction of the scattered light in the total intensity. Our results
show that for our five dusty clumps, the contribution of the
scattered light is less than 5\%, even in the MUV bands and that the $f$ and
$\Gamma(n)$ values changed by less than 5\%, if the contribution is accounted for. 

\subsubsection{Unresolved Reddening Structure in One \hst\ Resolution Element}
A dusty cloud likely has a fractal structure on all scales and extinction
variations should exist on scales smaller than one our \hst\ resolution element,
i.e. 0.13\arcsec\, $\sim$0.5 pc. This could bias our extinction
curve (see also, \S\ref{ss:f_a_uvot}). To examine this effect, 
we use the extinction map of the nearby Orion A and B molecular clouds, 
kindly provided by Jouni Kainulainen.~\citet{kai09} have used 
the NIR color excess
of background stars to map out the extinction distribution in the clouds. 
Because of their proximity (414$\pm$7
pc,~\citealt{men07}), the physical size of one pixel in 
these extinction map is 0.18 pc (1.5\arcmin ) and is roughly 
three times better than that of our \hst\ mosaics in the M31 bulge. By using the steps listed in Appendix D, we
find that the structure of the extinction variation in one \hst\ resolution element just
introduces less than 3\% and 2\% uncertainties into the $f$ and
$\Gamma(n)$ values. 

\section{Results}\label{s:result}
The extinction curves, expressed in terms of $A_n/A_{F547M}$, for the 
five dusty clumps are shown in Fig.~\ref{f:extin_slope}. 
We also provide the extinction curve averaged 
over the five dusty clumps. Although F275W and
UVW1 are from different detectors, the
similarity of the relative extinctions in these two bands is
apparent and demonstrates the reliability of our correction in
\S\ref{ss:f_a_uvot}. Because of the `Red Leak' problem, the effective
wavelength of the UVW1 is shifted to longer wavelengths, which
explains why the relative extinction in the UVW1 band is slightly smaller than
that in the F275W band. 

In Fig.~\ref{f:extin_slope}, we compare the extinction curves in the
five dusty clumps with those of the MW and
the MCs, as well as with the results of~\citet{mel00}. 
In the optical bands, our
extinction curves match the three empirical extinction curves 
well, while the relative extinction curve ($A_n/A_V$) of~\citet{mel00} 
is steeper than ours (see the middle right panel 
of Fig.~\ref{f:extin_slope}). Our relative extinctions in 
the two NIR bands (F110W and F160W) 
are larger than those of the MW and the MCs.
 But, because the absolute extinction
in the clumps is not large, giving little 
constraints on the extinction curve in the NIR, we do not include these
two bands in the subsequent quantitative analysis.  
Our extinction curve increases steeply in the
UV bands (F275W to UVW2), compared to that of the MW. The variations 
of $A_n/A_{F547M}$ among the five dusty 
clumps in the \swift/\uvot\ UV bands are substantial.
 However, the accuracy of $A_n/A_{F547M}$ in these three bands is limited by
the poor angular resolution and sensitivity of the \swift/\uvot.

We fit the extinction curves $\Gamma(n)$ in Fig.~\ref{f:extin_slope}
with~\citet{fit99} models, which parameterize the shape of the extinction
curve, including the possible 2175\AA\ bump. Because some of the
filters have the `Red Leak' problem, their effective wavelengths are
uncertain. We thus instead calculate the inferred extinction
curves (i.e. $\Gamma(n)_m$) that would be measured in the 
filters. We first use the best-fit stellar population model in the
unobscured regions~\citep{don13}, to calculate the spectral energy
distribution (SED) within each filter band pass. We then use the same
model to obtain the SED attenuated with the~\citet{fit99} models
 to calculate $\Gamma(n)_m$. We consider values of
1.5$<R_V<$4.5 (with a step size of 0.1, while the other parameters are
fixed) to minimize $\chi^2=\sum_{n}\frac{(\Gamma(n)-\Gamma(n)_m)^2}{(\delta
  \Gamma(n))^2}$, where the sum is over the filters from
UVW2 to F814W and the variable $\delta
  \Gamma(n)$ is the uncertainty of $\Gamma(n)$. The
final $R_V$ in the extinction model of~\citet{fit99}, 
ranging from 2.4-2.5 for the five dusty clumps, are listed in
Table.~\ref{t:f_a_uvot}. We use our Monte Carlo method to derive the
uncertainty of $R_V$ ($\delta R_V$). In
  Table.~\ref{t:f_a_uvot}, we also give the $R_V$s for the extinction model of~\citet{car89}, for
comparison with the result of the inner Galactic
Bulge of~\citet{nat13} in \S\ref{s:discussion}. 

The average $R_V$ values of the five dusty clumps with the extinction
model of~\citet{fit99} is 2.4$\pm$0.05, which are significantly 
smaller than the average Galactic $R_V$~\citep[3.1;][]{fit04}, and is even 
slightly steeper than the SMC
extinction curve~\citep[$R_V$=2.74$\pm$0.13;][]{gor03}. The
comparison between our observed $\Gamma$(n) and those predicted
by~\citet{fit99} with the corresponding $R_V$ is presented 
in Fig.~\ref{f:extin_model}. The model fits the observed extinction curves well in
the optical, with large deviations in the UV bands (from U to
UVW2). The clumps also show different degrees of agreement with
the~\citet{fit99} extinction curves. Clump A matches the model well,
whereas Clumps B and E agree only within their respective $\delta\Gamma(n)$ for the three
\swift/\uvot\ UV
filters, but show less overall agreement than Clump A. For Clumps
C and D, however, the observed
$\Gamma$(n) of the three \swift/\uvot\ filters are systematically
larger than those predicted by the model and especially,
for $\Gamma$(UVM2), by more than 2$\delta\Gamma(n)$ (Clump C,
2.4$\delta\Gamma(n)$ and Clump D, 3.0$\delta\Gamma(n)$). In spite of
these variations, the average 
extinction curve is well fit by the~\citet{fit99} model in most of the bands, although
a significant
enhancement in the UVM2 is apparent. 

\section{Discussion}\label{s:discussion}
The extinction curve is determined by
the size and compositions of dust, which could be related to many
factors, such as the metallicity of a molecular cloud and 
its environment. The metallicity alone is unlikely to be able to
explain the variations among the extinction curves in
the MW and the MCs. Different sightlines that have similar metallicity
gas can show very different extinction curves. Indeed, the extinction curves toward a few
lines of sight in the MCs are found to be similar to the Galactic extinction
curve~\citep{gor03}, whereas sightlines toward four stars in the Milky Way have steep
extinction curves that lack the 2175\AA\ bump~\citep{val04}. Therefore, 
factors other than 
metallicity must play important roles in determining the shape of
an extinction curves.~\citet{gor03} point out that the differences
between the MW and the MCs extinction curves may be due to their
sampling different environments. In particularly, most of the studied extinction curves in the MCs are from active
star-formation regions, where strong shocks and UV photons may conspire to destroy
large dust grains, whereas, those in the MW are typically toward runaway main-sequence OB
stars.  Within M31, the metallicity in the bulge is
comparable to regions of the disk~\citep{ros07}, for which~\citet{bia96} have
derived shallower extinction curves. Therefore, unless the clouds are
due to accreted low metallicity gas, it seems unlikely that
metallicity is responsible for those steep curves. 
 
We have found that the extinction curves in the CNR of M31 
are steep. 
We naively expected that the extinction curve there should be similar to or even flatter than the MW one,
because of their comparable metallicity and low star formation rate,
but found just the opposite. In fact, the CNR of M31 is
not the only galactic bulge with steep
extinction curves in the Local Group. The Galactic inner bulge has
recently been suggested to have a similar non-standard optical extinction curve~\citep{uda03,sum04,rev10,nat13}. These authors use red 
clump stars as standard candles, due to their nearly constant 
magnitude and color at high metallicities.  
They derive the foreground extinction in the optical and NIR bands
(V, I, J and K) from the differences of the observed and
intrinsic magnitudes/colors of the red clump stars toward
different sightlines in the Galactic Bulge. They find that the
relative exinction could not be explained by the standard Galactic extinction
curve ($R_V$=3.1), and must instead be steeper.~\citet{nat13} report
a value of $R_V$=2.5 toward the Galactic Bulge, with the extinction curve
model of~\citet{car89}, which is similar to the
$R_V$ value we have obtained in the CNR of M31. This consistency suggests that
a steep extinction curve could be common in galactic
bulges.

The extinction curves could be steepened by eliminating large
grains. It is possible that they have been destroyed
by interstellar shocks. Recombination lines (H$\alpha$, [N {\small \rm II}], [S {\small
  \rm II}], [O {\small \rm III}]) have been found in the CNR of M31
by~\citet{jac85} and arise from regions that are morphologically
similar to that of the dust emission~\citep{li09}. Therefore, 
the recombination lines are from the surfaces of the dusty molecular
clouds. Because the [N {\small \rm II}] line is
stronger than the H$\alpha$ line, these recombination lines are
suggested to be excited by shocks~\citep{rub71}. We speculate that the shocks from
supernova explosions or past activity of M31* have evaporated 
large dust grains and steeped the extinction curve. For example,
recently,~\citet{phi13} find that the interstellar medium of host galaxies
surrounding 32 Type Ia supernovae has $R_V<2.7$ with a mean value of
2.06. 

We also find the 2175\AA\ bump in our extinction curves, 
which is generally thought to
  be due to small graphite grains. The 2175\AA\ bump is especially
  strong in the extinction curve of Clump D, or probably Clump C. The former 
is located 30\arcsec\ (113 pc in projection) southeast of the
D395A/393/384 clump studied by~\citet{mel00}. This small and
compact clump core has a size $<$ 2 pc~\citep{li13} and appears 
dark in the \hst\ F275W, F336W and F390M images, consistent with its high $f$
value. The high metallicity of the clump may
provide the necessary carbon and silicon to construct small graphite
grains. Among the five dusty clumps, Clump D seems to have the smallest
median $A_{F547M}$, but the largest 2175\AA\ bump. The 2175\AA\ bump is weaker in the
extinction curve of Clump
B or E, which both have high $A_{F547M}$. 
This situation is probably reminiscent of the four lines of sight in
the MW~\citep{val04} through 
dense molecular clouds, which have weak 2175\AA\ bumps, 
for example, HD62542 ($A_V$=0.99$\pm$0.14) and
HD210121 ($A_V$=0.75$\pm$0.15). 

 Future \hst/STIS spectra in the mid-ultraviolet range are needed
  to confirm the potential 2175\AA\ bump in Clumps C and D. 
Because of the `Red Leak' problem and old stellar
  population in the M31 bulge, we need to assume the SED to
  compare the observed relative extinction in thirteen bands 
with the model. With the UV spectra, we can directly derive the
extinction curve, as well as the parameters of the 2175\AA\ bump,
  such as, its centroid and width. 

\section{Summary}\label{s:summary}
In this paper, we have presented the first study of 
the extinction curve within the central 1\arcmin\ region of M31 from
the MUV to the NIR. We have used \swift/\uvot\ and \hst\
\wfc3/\acs\ observations in thirteen bands 
to simultaneously constrain the line-of-sight locations and the relative extinction
$A_n/A_{F547M}$ of five dusty clumps in this region. Instead of 
fixed certain line-of-sight locations for these clumps, we have developed a 
method to determine their background stellar light fraction ($f$) directly from the
  observations. 

We have shown that the extinction curve is generally steep in the
circumnuclear region of M31, 
where the metallicity is super-solar. The derived $R_V$=2.4-2.5 is
similar to that found toward the Galactic Bulge. We discuss this
consistency which leads us to conclude that large dust grains are
destroyed in the harsh environments of the bulges, e.g., 
via potential shocks from supernova
explosions and/or past activities of M31*, as indicated by the strong
[N{\sc II}] recombination lines from 
the
dusty clumps. 

The extinction curves of the five dusty clumps show significant variations 
in the mid-ultraviolet. Some of the extinction
  curves can be explained by the extinction
  curve model of~\citet{fit99}. Others, most notably Clump D (probably
  also Clump C),  
shows an unusually strong 2175\AA\ bump, which is weak elsewhere 
in the M31 disk~\citep{bia96}. 



\section*{Acknowledgments}
We thank the anonymous referee for a thorough, detailed, and
constructurive commentary on our manuscript. We are grateful to Bruce Draine for many valuable
comments on the dust scattering. H. D. acknowledges the 
NASA support via the grant GO-12055
 provided by the Space Telescope Science Institute, which
is operated by the Association of Universities for Research in
Astronomy, Inc., under NASA contract NAS 5-26555. Z. L. acknowledges
support from NASA grant GO-12174 and NSFC grant 11133001.

\appendix
\section{Dependance of $A_{n}/A_{F547M}$ on Background Spectrum}
Compared with spectroscopic observations, broad-band photometric
studies of the extinction curve do have certain drawbacks. 
For a dusty clump with certain extinction curve, the effective
wavelength and hence the relative
extinction, $A_{n}/A_{F547M}$, of a filter could depends on 
the shape of the background spectrum. If there is more 
emission in the shorter wavelength (for example, due to the presence of a young stellar
population), the effective wavelength of the filter could then shift to the
shorter wavelength and the $A_{n}$ becames larger, and vice versa. 

We use the stellar synthesis model, Starburst99~\citep{vaz05}, to
examine the variation of the
relative extinction ($A_{n}/A_{F547M}$) for various incoming
spectra. Considering the high metallicity 
and no evidence for any recent star formation in the 
M31 bulge, we assume a background stellar population with
solar or super-solar metallicity (2.5$Z_{\odot}$) and age from
100 Myr to 10 Gyr. First, we redden the spectra of instantaneous
starbursts of different ages and/or metallicities by using the
extinction curves of the MW, LMC
or SMC and various absolute extinction E(B-V) (=$A_B-A_V$, hereafter
EBV for short). Second, we convolve this reddened spectra with the
transmission curves of the thirteen filters to derive the observed
magnitude (m$_{n}$(EBV)) by using `SYNPHOT' in `IRAF'. Third, we
fit the A$_{n}$(EBV) (=m$_{n}$(EBV)-m$_{n}$(0)) as a linear function of EBV 
to derive A$_{n}$/EBV, as well as $\frac{A_{n}}{A_{F547M}}$=$\frac{A_{n}}{EBV}$ / $\frac{A_{F547M}}{EBV}$. 


The ranges of A$_{n}$/$A_{F547M}$ of the MW and the MCs
for the instantaneous starbursts with solar metallicity are listed in
Table.~\ref{t:alambda_evo}. 
For the \hst\ filters, this ratio changes less than
6\%. On the other hand, for the three UV filters of \swift/\uvot , this ratio is
sensitive to the background stellar population 
and decreases by $\sim$45\%, $\sim$17\%,  $\sim$25\%
for UVW2, UVM2 and UVW1 from 100 Myr to 10 Gyr. The `Red Leak'
problem of UVW2 and UVW1 (see \S\ref{ss:swift_obs}) is the main reason
to explain the large variation of A$_{n}$/$A_{F547M}$ at these two
bands. We also compare A$_{n}$/$A_{F547M}$ between the $Z_{\odot}$ and
2.5$Z_{\odot}$. Since the metallicity can significantly change the UV
colors of the underlying stellar population, the systematic shift of A$_{n}$/$A_{F547M}$
between the $Z_{\odot}$ and 2.5$Z_{\odot}$ could reach $\sim$13\% (UVW2
and UVM2) or 5.5\% (UVW1) for the old stellar population, assuming the
SMC extinction curve, which rises 
quickly in the UV bands. Instead, the change of A$_{n}$/$A_{F547M}$ for the nine \hst\
filters is less than 1\% .  

Considering the dependence of A$_{n}$/$A_{F547M}$ on the background 
spectrum, in \S\ref{s:result}, we adopt the
underlying stellar populations characterized by~\citet{don13}. With
the same \hst\ dataset,~\citet{don13} divide the  
extinction-free region in the southeast of
the M31 (see also Fig.~\ref{f:ratio}) into concentric annuli of 
5\arcsec\ wide and fit the SED in
each annulus individually with two 
metal rich instantaneous starbursts (intermediate-age: $\sim$700 Myr,
$\sim$2Z$_{\odot}$ and old: $\sim$12 Gry, $\sim$1.3Z$_{\odot}$). 
They provide the radial profiles of ages and metallicities of these two
stellar populations along the minor axis of the M31 bulge, which we
use to construct the intrinsic spectra of the background stellar
population for our five dusty clumps. Then, we use the same method
above to derive the model A$_{n}$/$A_{F547M}$, used to compared with the observed relative
extinction in \S\ref{s:result}. For $R_V$=2.5, we calculate the 
standard deviation of  $A_n/A_{F547M}$ among the annuli between 10\arcsec\ to
60\arcsec , which are 4.5\%, 1.2\%, 0.9\% in the UVW2, UVM2 and UVW1
band, smaller than the observed
uncertainties of these filters in \S\ref{ss:f_a_uvot}.

\section{Flattening Bias}
We employ the method of~\citet{fri11} to remove the `flattening bias'. The
method is based on the following assumption: the intrinsic starlight and extinction
distribution in the three UV filters could be
mimicked from the other observations, such as the \hst\ F336W
image. Although the F275W band is closer to the three UV bands, it has
larger photometric uncertainty, compared to the F336W band 
(see Table.~\ref{t:obs}). 

 We use the following steps to constrain the relative extinction of
 the \swift/\uvot\ filters in the CNR of
 M31. We assume that $S_{F336W}$ derived
in \S\ref{ss:intri} is similar to the intrinsic light
distribution of the three
UV filters, the extinction distribution of which
 is also proportional to that of the
F336W band, i.e. $A_n$ = $\Gamma(n,F336W)\times A_{F336W}$. For the 
observed and intrinsic F336W images,
we first set
the `source' pixels to 0 and then convolve the images with the 
\swift/\uvot\ PSF (\S\ref{ss:swift_obs}). The new observed and
intrinsic F336W images are used
to derive $A(F336W)$ through the Eqn~\ref{e:ratio} ($f$ from
\S\ref{ss:f_a}). Second, for a certain $\Gamma(n,F336W)$, we 
obtain a extinction map, $A_n$, which is used to redden the 
intrinsic F336W images. Third, for each \swift/\uvot\ pixels,
we calculate the ratio of the sum of the reddened and 
intrinsic intensities of the corresponding \hst\ pixels above. Fourth, we
use `MPFIT' to search for $\Gamma(n,F336)$ that best matches the
above model ratio with the one between observed and intrinsic
intensities of dusty pixels in \swift/\uvot\ images. The median and
standard deviation of $\Gamma$(n,F336W) for each dusty clump
 are listed in Table~\ref{t:uv_rela}.

\section{Effect of the Scattered Light}
We estimate the effect of the scattered light on our derived extinction curve. 
A derpojected S\'ersic model, assumed to be sphereically
symmetric, is adopted to approximate the radiation field in the M31
bulge. For a dusty clump inside the bulge, we derived
$I_s$ (the scattered intensity), $I_{ua}$ (the total starlight
intensity from the front side) and $I_a$ (from the back side). 
The ratio, $\frac{I_s}{I_{ua}+I_a}$, represents
the importance of the scattered intensity in Eqn.~\ref{e:ratio}.~\citet{gro12} suggested
that the dust is probably optical thin even in the near-UV
wavelength. This is supported by our data: $N_H$ never significantly
exceeds $10^{21}$ {\rm H cm}$^{-2}$~\citep{li13} and the optical depth should
be smaller than 0.6 even in the UVM2 band. Therefore, for simplicity,
we neglect multiple-scattering events. 

We constructed the deprojected S\'ersic model for the stellar
light emissivity, $\nu(r)$ 
(in units of ergs $s^{-1}$ $cm^{-3}$, where r is the physical radius,
in units of pc)  by using Eqn. 20a in~\citet{bae11}. According to~\citet{li13},
the average S\'ersic index of the M31 bulge 
for the ten \hst\ filters is $\sim$2.2,
which is consistent with that obtained by~\citet{kor99}. The effective
radius of the S\'ersic model is 313 pc. 
We calculated $\nu(r)$ for the central 1 kpc$\times$1 kpc
(i.e. $\sim$250\arcsec$\times$250\arcsec\ in projection) with a spatial resolution of
L=4 pc. This region contributes $\sim$98\% of the total starlight in the
M31 bulge.  

Then, we put a dusty clump into the model with different sky coordinate ($R$,
$z$) to derive $I_s$, $I_{ua}$, $I_a$, and then $\frac{I_s}{I_{ua}+I_a}$. 
`$R$' is the projected distance and `$z$' is the line of sight
distance, in units 
of pc. Both the `$R$' and `$z$' are centered at the nucleus. 
`$z$' is the monofonic function of $f$. 
Since we knew the
$\nu(r)$, for each `z', we could derive the relative $f$. We adopted the
extinction cross section per hydrogen ($\sigma_{ext}(n)$, cm$^2$/H) and differential scattering
cross section per hydrogen ($\frac{d
  \sigma_{sca}(n,\theta)}{d\Omega}$, cm$^2$ H$^{-1}$ sr$^{-1}$) 
from~\citet{dra03}. We wrote
the following equations: 
\begin{eqnarray}
I_{ua}\propto\sum_{z' > z(f)} L^3\nu (r')\\
I_{a}\propto\sum_{z' < z(f) } L^3\nu (r') \times 10^{-0.4\times N_H\times
  \sigma_{ext} (n)}\\
I_s\propto\sum_{r'} \frac{L^3\nu (r')\times L^2}{4\pi\times |r'-r|^2}\times N_H\times\frac{d
  \sigma_{sca}(n,\theta)}{d\Omega}
\end{eqnarray}
$L^3\nu (r')$ is the stellar emission for a block at $r'$. $\frac{L^2}{4\pi\times |r'-r|^2}$ is the solid
angle of the dusty clump (`r') toward the stellar intensity of the
block at `$r'$'. 

Through the above model, we derive $\frac{I_s}{I_{ua}+I_a}$ for our thirteen filters with
different projected distances (`$R$' from 0 to 250 pc, i.e. 1.1\arcmin
), $f$ (0.2 to 0.9, with a binsize of 0.1) and absolute extinction
($A_V$, 0.1 to 0.7, with a bin size of 0.2) for MW-type
dust~\citep{dra03}. Since $\sigma_{ext}(n)$ and $\frac{d
  \sigma_{sca}(n,\theta)}{d\Omega}$ of the SMC-type dust is
smaller than those of the MW-type dust by nearly an order of magnitude (see Fig. 3
and Fig. 4 of ~\citealt{dra03}), the effect of scattering for a
SMC-type dust is much smaller than that of MW-type dust. In
Fig.~\ref{f:scat_pub}, we present $\frac{I_s}{I_{ua}+I_a}$ as a
function of wavelength for our
thirteen filters at two projected distances (100 pc and 250 pc) with
four different $f$ value (0.3, 0.5, 0.7 and 0.9). The
$\frac{I_s}{I_{ua}+I_a}$ is larger in the shorter wavelength, because
of the large extinction and scattering cross section. Since the
photons prefer forward-scattering~\citep{dra03}, the more
starlight behind the dusty clump (larger $f$ value), the more
starlight will be scattered into the line of sight. When the clumpy
is far away from the M31* (larger projected distance, `$R$'), the
$I_{ua}$+$I_a$ decreases and more
starlight from $r < R $ could be reflected to the lines-of-sight of observers.

We then used the above information to estimate the contribution of the
scattered light in our five clumps. The averaged projected distances
and $f$ of the five clumps, derived by assuming that the scattered
emission is negilible, are listed in Table.~\ref{t:f_a_uvot}. Since 
our dusty clumps have either small projected
distances (Clump A, B and C), small $f$ (Clump E) or low absolute extinction ($A_{F547M}$,
Clump D), we expected that the $\frac{I_s}{I_{ua}+I_a}$ should be roughly less
than 5\% in all our thirteen filters. We further quantified the
effect of the scattered light on our results. From the averaged
projected distance, $f$ and median $A_{F547M}$, we derived
$\frac{I_s}{I_{ua}+I_a}$ of the thirteen filters 
for each clump. Then, we divided $\Re$(n,k) in
Eqn.~\ref{e:ratio} by (1+$\frac{I_s}{I_{ua}+I_a}$) to
exclude the scattered emission. After reprocessing the steps 
in \S\ref{ss:f_a} and
\S\ref{ss:f_a_uvot}, we found that the $f$ for the five dusty
clumps decreases by less than 5\% and the $\Gamma(n)$ changes
  less than 3\% in the UV bands. Therefore, we concluded that the
ignorance of the scattered emission in Eqn.~\ref{e:ratio} has little
effect on our result. 


\section{Effect of Unresolved Reddening Structure in One \hst\ Resolution Element }
We use the extinction map of Orion molecular clouds from~\citet{kai09} to measure the
bias introduced by unresolved dust structure in one \hst\ resolution
element. We first interpolate the intrinsic light
distribution ($S_n$) at the F547M band in \S\ref{ss:intri} into a finer pixel size
(0.048\arcsec , $\sim$0.18 pc in the M31 bulge). Then, 
by using the extinction map from~\citet{kai09} and the $f$ values listed in
Table~\ref{t:f_a_uvot}, we simulate the observed light distribution ($I_n$) in
the five dusty clumps in Fig.~\ref{f:ratio} with Eqn.~\ref{e:ratio},
which is then downgraded into the original pixel 
size of our \hst\ images (0.13\arcsec ). $A_V$ in the Orion molecular
cloud reaches 25 mag, much larger than those of our five dusty
clumps. Thus we first scale the absolute
extinction of the Orion molecular cloud to match the minimum 
$\Re_{F547M}=\frac{I_{F547M}}{S_{F547M}}$ 
of the simulated and real (\S\ref{ss:hst_obs}) observed light
distributions for each clump. We then estimate the uncertainty of the
simulated observed light in individual pixels, from dispersion
 $\sigma_{n}$ (see \S\ref{ss:hst_obs}). We finally use the average relative
extinction, $\Gamma(n)$, listed in Table~\ref{t:f_a_uvot}, or the
MW/SMC/LMC-type extinction curves, to get the
$A_n$ distribution at the other nine bands for each dusty
clump and produce the simulated observed light distribution at these
bands, with the same method above. Then, the
similar steps used in \S\ref{ss:mole} and \S\ref{ss:f_a} are used to
calculate the $f$ and $\Gamma(n)$ for these five dusty clumps. 
We find that the differences between the input and output
values of $f$ and $\Gamma(n)$ are less than 3\% and 2\%,
respectively.  This means that thanks to the high resolution of our \hst\
observations, we may resolve enough detail structures of the dusty
clumps in M31 bulge. Therefore, we can neglect this effect due 
to the subpixel dust structure.

\begin{figure*}[!thb]
  \centerline{
       \epsfig{figure=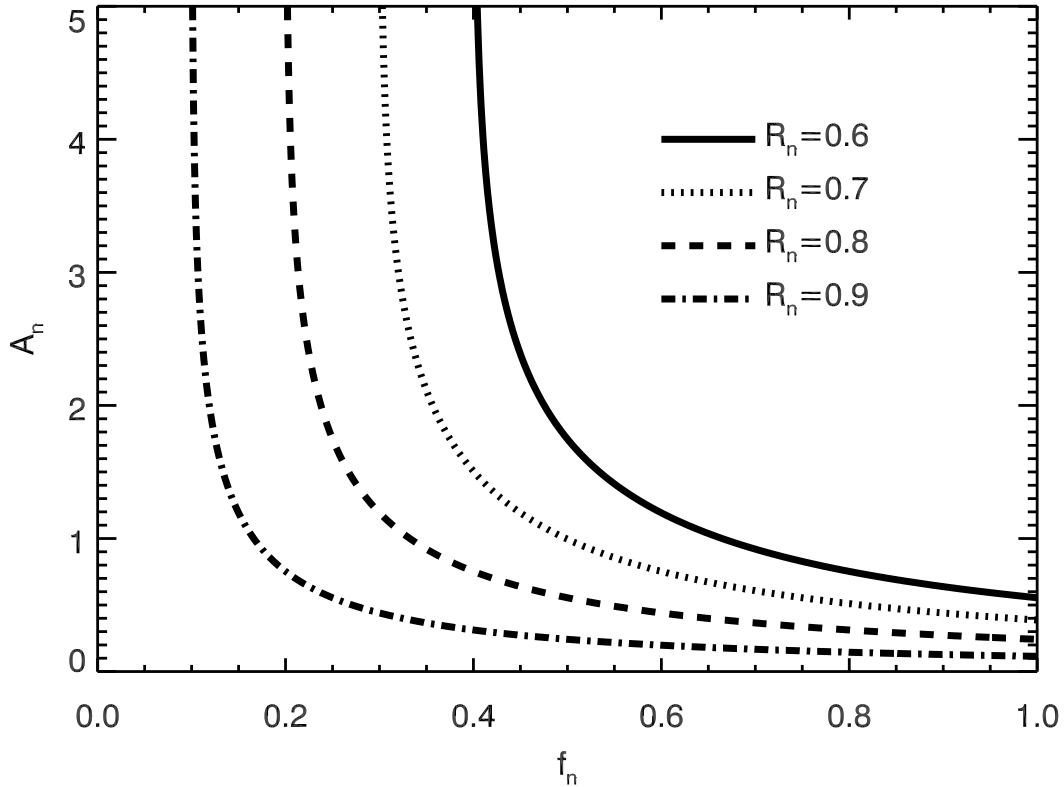,width=1.0\textwidth,angle=0}
      }
 \caption{ The relationship between the fraction of obscured
   starlight ($f_n$) and the absolute extinction $A_n$ for different 
$\Re_n$=$I_n/S_n$ derived from Eqn.~\ref{e:ratio} ($I_n$ and $S_n$ are
  the {\sl observed} and {\sl intrinsic} intensities). $f$ is
  anti-correlated with $A_n$. When $f_n$
  decreases, (i.e., more stars in front of a dusty clump), $A_n$
  increases for the same $\Re_n$.}
 \label{f:f_a_rela}
 \end{figure*}

\begin{figure*}[!thb]
  \centerline{
       \epsfig{figure=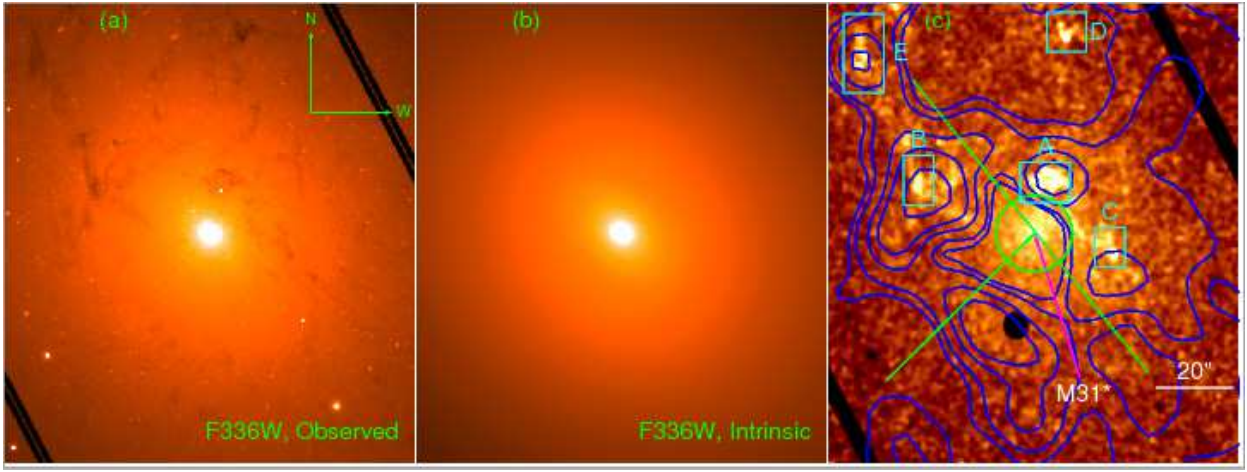,width=1\textwidth,angle=0}
}
\caption{Left: \hst/\wfc3\ F336W intensity map of the central
  2\arcmin$\times$2\arcmin\ of the M31 bulge. Middle: The intrinsic
  light distribution of \hst/\wfc3\ F336W derived by the method in
  \S\ref{ss:intri}. Right: The intensity
  ratio map between the F160W and F336W bands,
 convolved with a Gaussian kernel of 9 pixels to reduce the
  noise. The five cyan boxes outline 
the dusty clumps used to constrain the extinction
curves (see \S\ref{ss:f_a}). The green circle shows the
central 10\arcsec\ around the M31*. The black spot to the southeast 
is the `death
star' (very low sensitivity) feature in the \wfc3 /IR detector.
The two black strips are excuded regions, which were covered by only one
 dithered exposure in the F275W or F336W bands and do not allow for
 cosmic-ray removal. The dashed and solid green lines represent the major and
 minor axes of the M31 bulge. The blue contours are from the Spitzer/IRAC
 `dust-only' 8 $\mu$m intensity map of~\citet{li09} }
\label{f:ratio}
\end{figure*}

\begin{figure*}[!thb]
  \centerline{
       \epsfig{figure=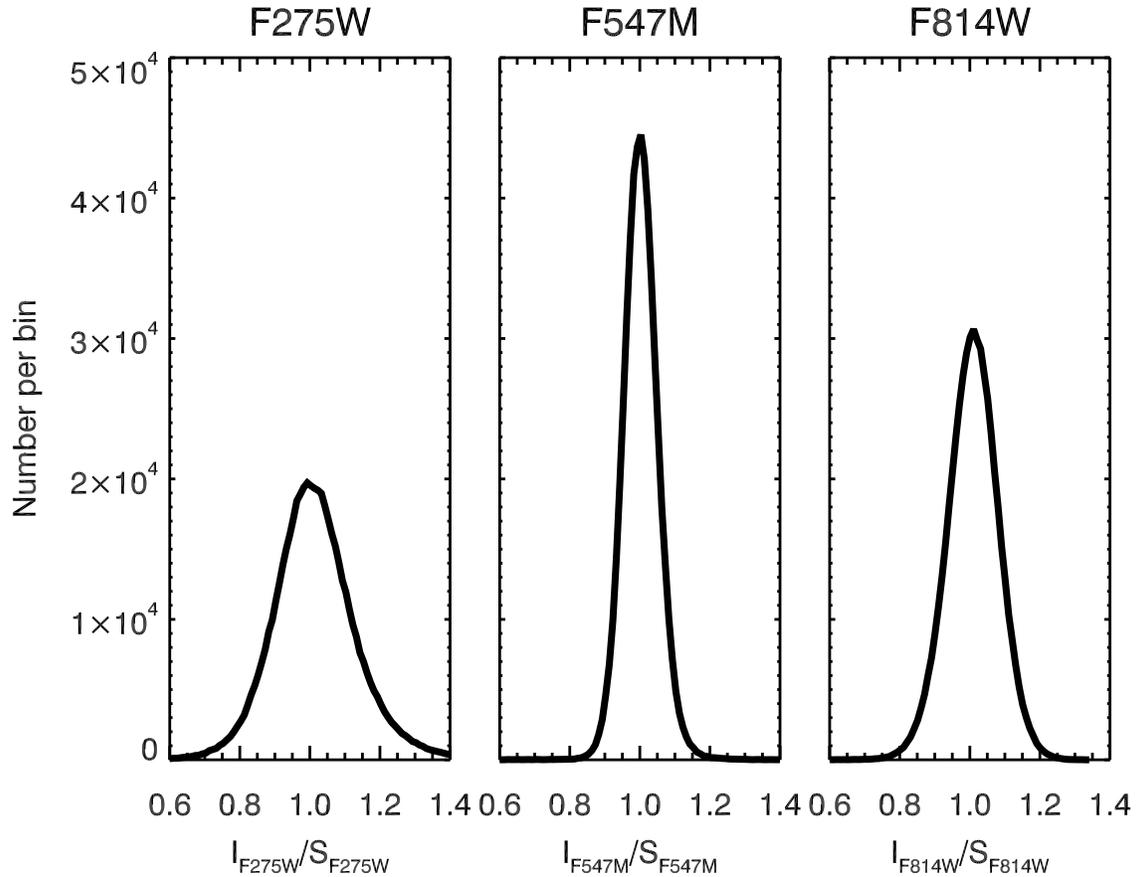,width=1\textwidth,angle=0}
}
\caption{The distribution of I$_n$/S$_n$ with a bin size of
  0.01 for pixels in the central
  2\arcmin$\times$2\arcmin\ of the M31 bulge at F275W, F547M and
  F814W. The curves can be fitted well by Gaussian functions, the
  centroid and standard deviation of which are $\sim$1 and 0.086/0.035/0.058 at
  F275W/F547M/F814W bands. The dispersions of the Gaussian functions can be
  explained by the photometric uncertainties at these three bands
  ($\sigma_n$, see Table.~\ref{t:obs}).}
\label{f:IvsS}
\end{figure*}

\begin{figure*}[!thb]
  \centerline{
       \epsfig{figure=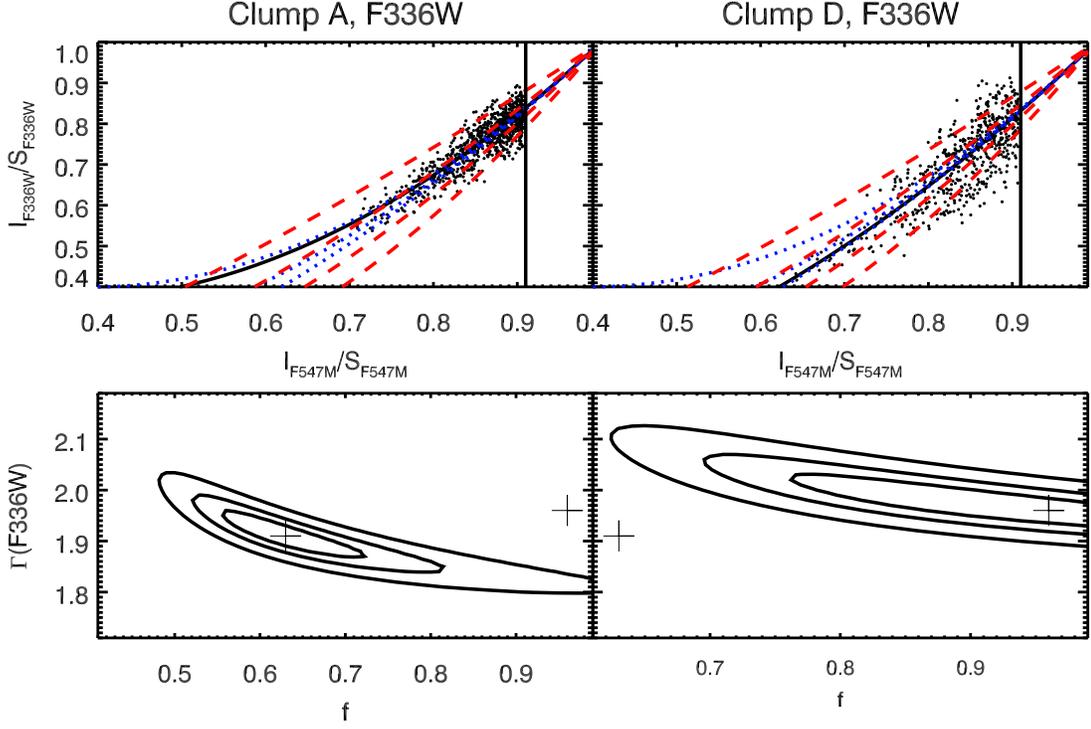,width=1\textwidth,angle=0}
       }
 \caption{Top panels: $I_{F336W}/S_{F336W}$ versus $I_{F547M}/S_{F547M}$ for Clumps A (left) and D (right). The `dots' represent the
   data points of individual `dusty' pixels of the clumps. The cutoff
   for $I_{F547M}/S_{F547M}$ around 0.9 is due to the criteria for
   selecting the `dusty' pixels (vertical black lines). The solid lines represent the best-fit
   relationship (Eqn.~\ref{e:f_a}, Clump A: $f$=0.63 and
   $\Gamma(F336W)$=1.91 and Clump D: $f$=0.96 and
   $\Gamma(F336W)$=1.96). 
The blue dotted lines represent the relationships with the 
best-fit $\Gamma(F336W)$ and $f$ fixed at 0.6 (bottom),0.8 (middle)
and 1.0 (top), whereas the red dashed lines represent 
the best-fit relationships with $f$ and $\Gamma(F336W)$ 
fixed at [0.7 (top),0.9 (middle-up),1.1 (middle-low),1.3 (bottom)]
$\times$ its best-fit value. For the `dusty' pixels with 
$I_{F547M}/S_{F547M}>0.85$, the $I_{F336W}/S_{F336W}$ is not
sensitive to $f$, but can constrain the $\Gamma(F336W)$. On the
other hand, the observed data points with $I_{F547M}/S_{F547M}<0.8$ can
 distinguish the different values of $f$. Bottom panels: $f$
versus $\Gamma(F336W)$ for Clumps A (left) and D (right). The
contours represent 60, 90 and 95\% confidence
levels. The `plus' symbols mark the best-fit $f$ and $\Gamma(F336W)$ for these
two Clumps.}
 \label{f:demonstrate}
 \end{figure*}

\begin{figure*}[!thb]
  \centerline{
       \epsfig{figure=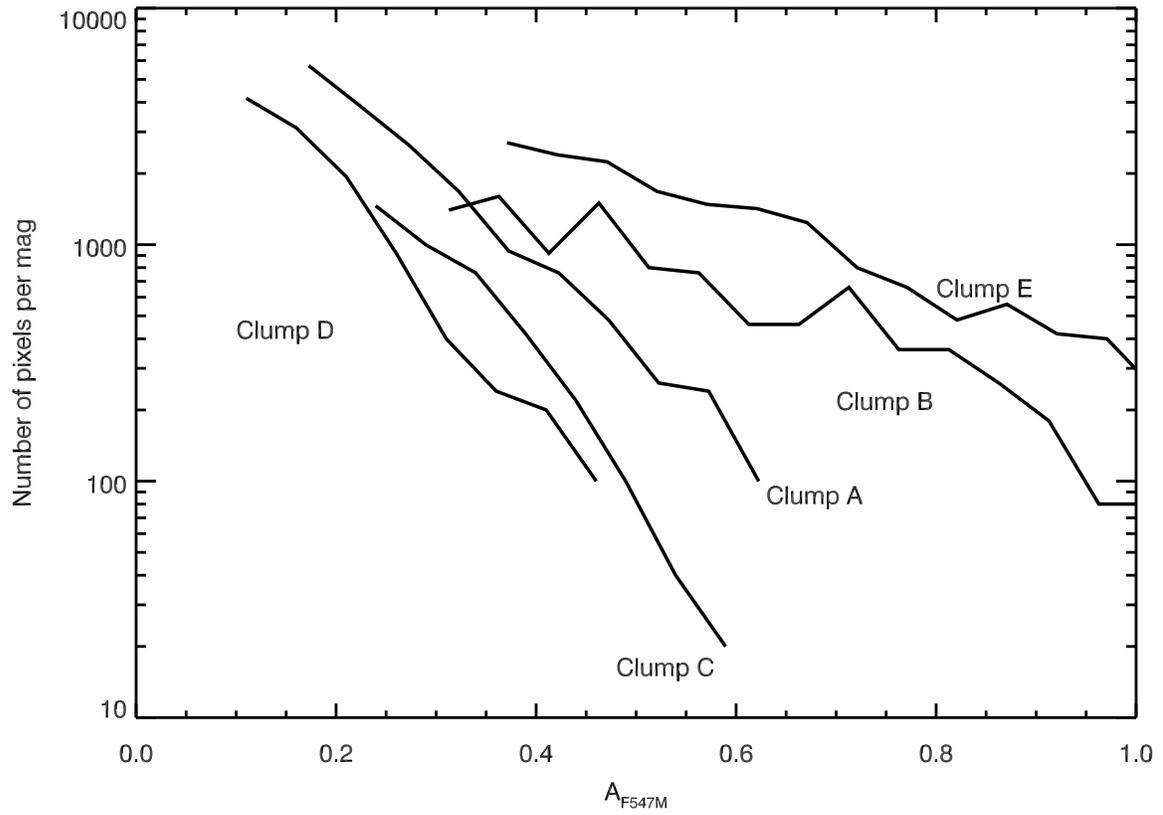,width=1\textwidth,angle=0}
       }
 \caption{The $A_{F547M}$ distribution of the `dusty' pixels for
   the five clumps (see \S~\ref{ss:f_a}). The bin size of the curves is 0.1 mag. }
\label{f:nh}
 \end{figure*}


\begin{figure*}[!thb]
  \centerline{
       \epsfig{figure=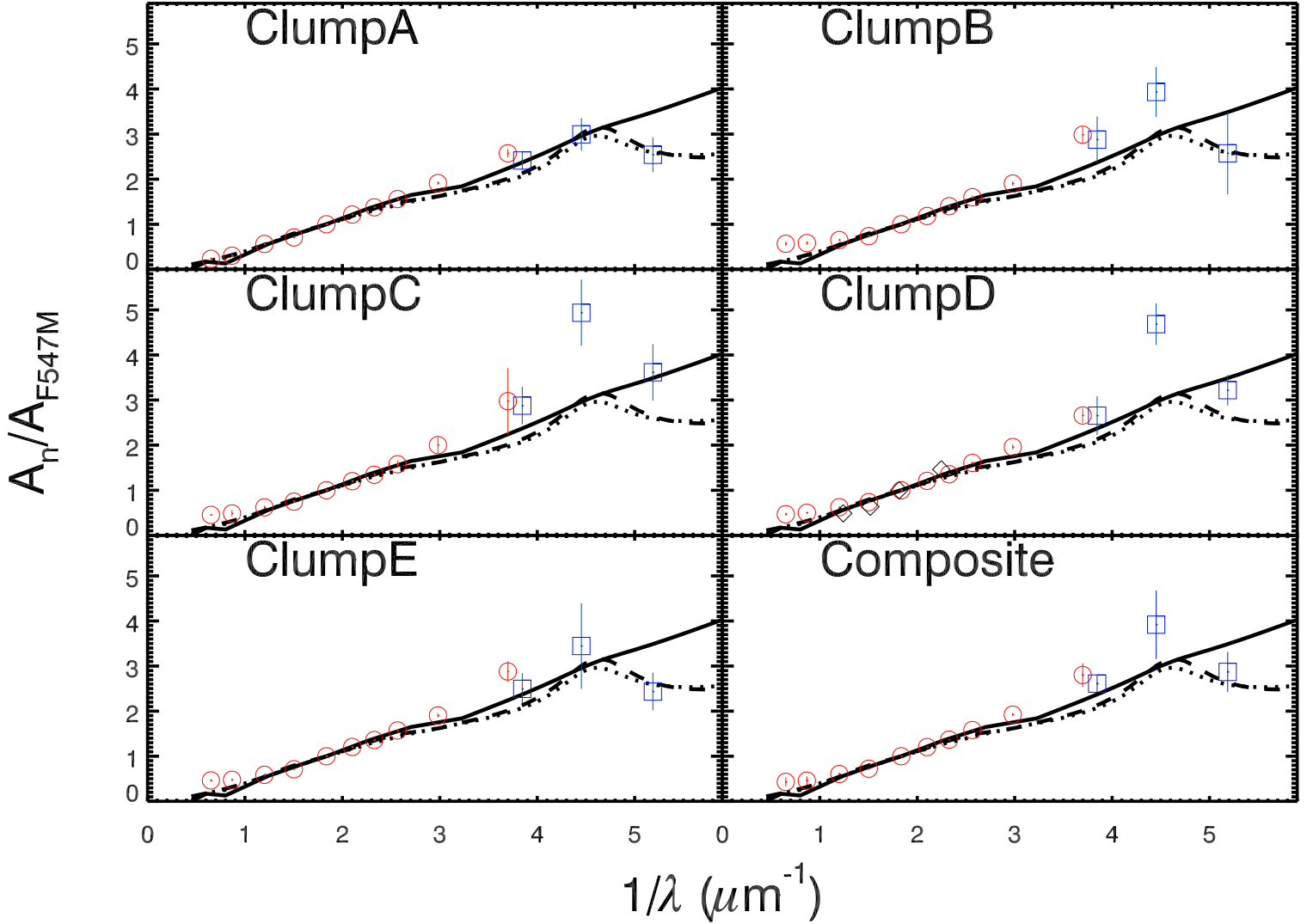,width=1.3\textwidth,angle=0}
       }
 \caption{The relative extinction ($A_n/A_{F547M}$) of the
   five dusty clumps and the averaged
   one (the low-right panel). The red `circle' and blue `box' symbols represent
   \hst\ and \swift\ filters, respectively. For comparison, the extinction curves of
   the SMC (solid),
   LMC (dotted)
   and MW (dashed) are also shown. The black `diamonds' in the middle-right
   panel represent the $A_{n}/A_V$ for the M31 bulge in~\citet{mel00}. 
    In the UV bands,
   1/$\lambda$ ($\mu$m$^{-1}$) $>$ 3.5, the extinction
   curves of our dusty clumps are steeper than that of the MW, even
   that of the SMC. Clumps C and D have large relatively
   extinction around 2175\AA\ (1/$\lambda\sim4.6$).}
\label{f:extin_slope}
 \end{figure*}

\begin{figure*}[!thb]
  \centerline{
       \epsfig{figure=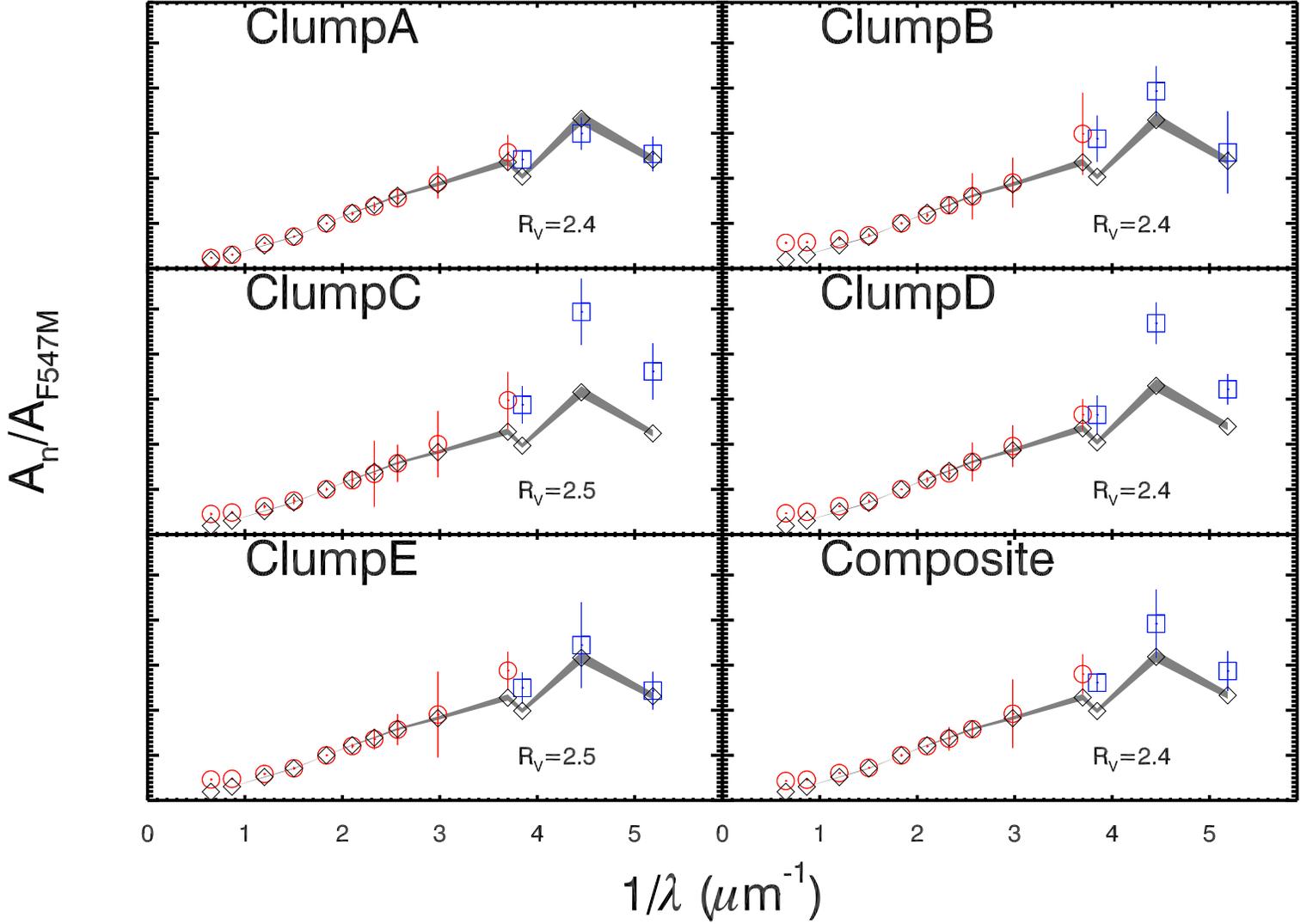,width=1.3\textwidth,angle=0}
       }
 \caption{The same as Fig. 6,  but 
compared with the $A_n/A_{F547M}$ (diamonds) predicted by the
   Galactic extinction model of~\citet{fit99} with the corresponding best-fit
   $R_V$ obtained in \S\ref{s:discussion}. The shaded regions indicate
   the range of $A_n/A_{F547M}$ with $R_V$ equal to the best-fit
   $R_V\pm0.1$. The $R_V$ values are listed in the
   lower right corners  of the panels. The extinction curves in the
   central 200 pc of
   the M31 bulge ($R_V$=2.4-2.5) are similar to the recent measurement
   toward the inner Galactic Bulge ($R_V$=2.5), which means that a steep
   extinction curve in the galactic bulge could be common.}
 \label{f:extin_model}
 \end{figure*}

\begin{figure*}[!thb]
  \centerline{
       \epsfig{figure=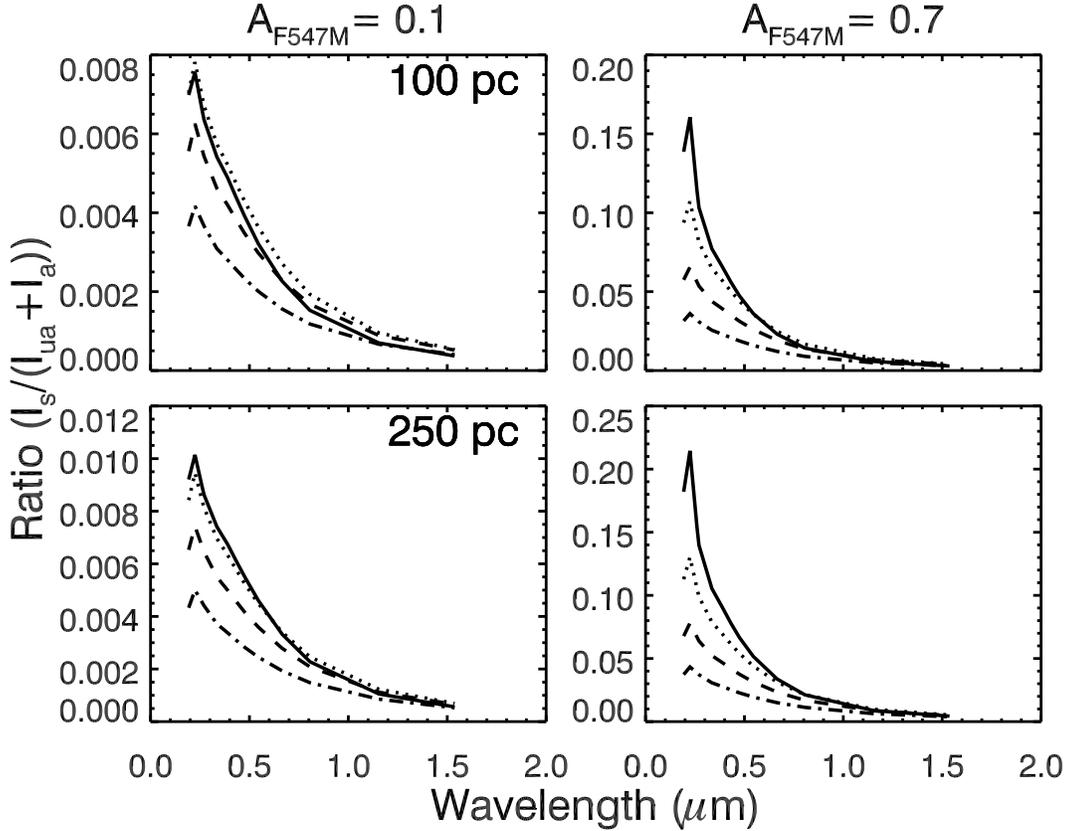,width=1\textwidth,angle=0}
       }
 \caption{Illustration of the intensity ratio $\frac{I_s}{I_{ua}+I_a}$, as a function
   of the wavelength, the projected distance of a clump to the M31*,
   the fraction of obscured starlight ($f$) and absolute extinction
   ($A_{F547M}$). $I_s$, $I_{ua}$ and $I_a$ are the scattered light, the
   starlight in front of the dusty clumps and the obscured
   starlight. The top panels are for the projected distances 100
   pc, while the bottom panel for 250 pc. The left column has
   $A_{F547M}$=0.1, while the right column has $A_{F547M}$=0.7. 
The lines in the plots from top to the bottom are
   for $f$ = 0.9 (solid), 0.7 (dotted), 0.5 (dashed) and 0.3
   (dash dot). The contribution of the scattered light,
   $\frac{I_s}{I_{ua}+I_a}$ decreases, following the increase of
   wavelength and the decrease of the projected
   distance/$f$/$A_{F547M}$. Considering the $A_{F547M}$, projected distances
   and $f$ of our five dusty clumps, even at the three Swift/UVOT UV
   bands, the contribution of the scattered light is less than 5\% .}
 \label{f:scat_pub}
 \end{figure*}

\begin{deluxetable}{cccccc}
  \tabletypesize{\small}
 \tablecolumns{6}
  \tablecaption{Multi-Wavelength Filters}
  \tablewidth{0pt}
  \tablehead{
  \colhead{Filter} & 
  \colhead{Detector} &
  \colhead{Central Wavelength (\AA )} &
  \colhead{Width (\AA )} & 
  \colhead{Exposure} &
  \colhead{$<\sigma_n/I_n>^a$} \\
  }
  \startdata
UVW2 & \swift /\uvot & 1928 & 657 & 106 ks & \\
UVM2 & \swift /\uvot & 2246 & 498 & 46 ks & \\
UVW1 & \swift /\uvot & 2600 & 693 & 153 ks & \\
F275W & \wfc3 /UVIS & 2704 & 398 & 925 s & 8.6\%\\
F336W & \wfc3 /UVIS & 3355 & 511 & 1250 s & 2.7\% \\
F390M & \wfc3 /UVIS & 3897 & 204 & 2700 s & 2.8\%\\
F435W & \acs /WFC & 4319 & 1038 & 2200-4600 s & 2.7\%\\
F475W & \acs /WFC & 4747 & 1458 & 1900 s & 3.0\%\\
F547M & \wfc3 /UVIS &  5447 & 650 & 2700 s & 3.6\%\\
F665N & \wfc3 /UVIS &  6656 & 131 & 2700 s & 4.4\%\\
F814W & \acs /WFC & 8057 & 2511 & 1715 s & 5.6\%\\
F110W & \wfc3 /IR & 11534 &  4430 & 699 s & 9.3\%\\
F160W & \wfc3 /IR &  15369 & 2683 & 1600 s & 11.1\%\\
\enddata
\label{t:obs}
\tablecomments{ a) The median ratio of the empirical noise to observed
  intensity of the ten \hst\ filters in the central
  2\arcmin$\times$2\arcmin\ ($\sim$450pc$\times$450pc) of M31 (see \S~\ref{ss:hst_obs}).}
\end{deluxetable}

\begin{deluxetable}{cccccc}
  \tabletypesize{\large}
  \tablecolumns{4}
  \tablecaption{Five Dusty Clumps}
 \tablewidth{0pt}
  \tablehead{
  \colhead{Name} & 
  \colhead{R.A. (degree)} &
  \colhead{Dec. (degree)} &
  \colhead{size} \\
}
\startdata
Clump A & 10.683603 & 41.272742 & 12.8\arcsec$\times$10.3\arcsec\ \\
Clump B & 10.695704 & 41.272921 & 7.7\arcsec$\times$12.8\arcsec\ \\
Clump C & 10.677435 & 41.268105 & 7.7\arcsec$\times$10.3\arcsec\ \\
Clump D & 10.681562 & 41.283549 & 10.3\arcsec$\times$10.3\arcsec\ \\
Clump E & 10.700926 & 41.282015 & 10.3\arcsec$\times$20.5\arcsec\ \\
\enddata
\label{t:clumps}
\end{deluxetable}
\begin{deluxetable}{c|c|c|c|c|c|c}
  \tabletypesize{\small}
 \tablecolumns{7}
  \tablecaption{Properties of the Five Dusty Clumps}
  \tablewidth{0pt}
  \tablehead{
  \colhead{} & 
  \colhead{Clump A} & 
  \colhead{Clump B} &
  \colhead{Clump C}&
   \colhead{Clump D} & 
  \colhead{Clump E} &
  \colhead{Average} 
  }
  \startdata
f &
0.63$\pm$0.066&0.37$\pm$0.020&0.47$\pm$0.133&0.96$\pm$0.201&0.32$\pm$0.014
& \\
$\Gamma(UVW2)^a$ & 2.54$\pm$0.387&2.57$\pm$0.913&3.62$\pm$0.627&3.22$\pm$0.340&2.43$\pm$0.419&2.87$\pm$0.380\\
$\Gamma(UVM2)$ & 2.99$\pm$0.360&3.93$\pm$0.554&4.94$\pm$0.736&4.68$\pm$0.462&3.45$\pm$0.952&3.92$\pm$0.502\\
$\Gamma(UVW1)$ & 2.41$\pm$0.211&2.88$\pm$0.512&2.88$\pm$0.414&2.65$\pm$0.429&2.50$\pm$0.344&2.61$\pm$0.091\\
$\Gamma(F275W)$ & 2.57$\pm$0.091&2.99$\pm$0.176&2.98$\pm$0.731&2.66$\pm$0.163&2.88$\pm$0.228&2.80$\pm$0.261\\
$\Gamma(F336W)$ & 1.91$\pm$0.038&1.90$\pm$0.053&2.00$\pm$0.180&1.96$\pm$0.058&1.91$\pm$0.049&1.92$\pm$0.029\\
$\Gamma(F390M)$ & 1.56$\pm$0.024&1.60$\pm$0.038&1.58$\pm$0.079&1.61$\pm$0.034&1.57$\pm$0.031&1.58$\pm$0.019\\
$\Gamma(F435W)$ & 1.38$\pm$0.018&1.40$\pm$0.028&1.34$\pm$0.042&1.35$\pm$0.023&1.36$\pm$0.023&1.37$\pm$0.018\\
$\Gamma(F475W)$ & 1.22$\pm$0.015&1.18$\pm$0.019&1.20$\pm$0.033&1.21$\pm$0.018&1.21$\pm$0.019&1.21$\pm$0.006\\
$\Gamma(F665N)$ & 0.71$\pm$0.016&0.74$\pm$0.025&0.75$\pm$0.031&0.74$\pm$0.017&0.71$\pm$0.016&0.73$\pm$0.013\\
$\Gamma(F814W)$ & 0.56$\pm$0.021&0.65$\pm$0.026&0.62$\pm$0.037&0.62$\pm$0.020&0.59$\pm$0.019&0.61$\pm$0.031\\
$\Gamma(F110W)$ & 0.30$\pm$0.027&0.58$\pm$0.041&0.49$\pm$0.075&0.50$\pm$0.025&0.48$\pm$0.027&0.46$\pm$0.094\\
$\Gamma(F160W)$ &
0.23$\pm$0.029&0.57$\pm$0.046&0.45$\pm$0.015&0.47$\pm$0.030&0.46$\pm$0.030&0.43$\pm$0.099\\
$\langle r\rangle^b$ & 65 pc & 121 pc & 80 pc & 232 pc & 235 pc
& \\
$\langle A_{F547M}\rangle$ & 0.25$\pm$0.1 & 0.5$\pm$0.18 &
0.31$\pm$0.08 & 0.18$\pm$0.07 & 0.56$\pm$0.20 & \\
$R_{V, Fit}^c$ & 2.4$\pm$0.06 & 2.4$\pm$0.08 & 2.5$\pm$0.13 & 2.4$\pm$0.07 &
2.5$\pm$0.07 & 2.4$\pm$0.05 \\
$R_{V, Car}^d$ &2.3$\pm$0.05 & 2.3$\pm$0.08 & 2.2$\pm$0.32 & 2.2$\pm$0.07 &
2.3$\pm$0.07 &2.3$\pm$0.05 \\
Number of \hst\ pixels & 837 & 499 & 202 & 555 & 863 & \\
$\chi^2$/d.o.f (d.o.f =9) & 2.7 & 3.4 & 2.6 & 4.4 & 1.9 & 4.2\\
\enddata
\label{t:f_a_uvot}
\tablecomments{a) $\Gamma(n)=\frac{A_n}{A_{F547M}}$, the relative extinction. b) `$\langle r\rangle$' is the averaged projected distance of each
clump to the M31*. c) The $R_V$ values for the extinction curve model of~\citet{fit99}. d) The $R_V$ values for the extinction curve model of~\citet{car89}.}
\end{deluxetable}

\begin{deluxetable}{c|c|c|c|c|c}
  \tabletypesize{\small}
 \tablecolumns{5}
  \tablecaption{The Extinction Relative to the F336W Band for the \swift/\uvot\ Filters}
  \tablewidth{0pt}
  \tablehead{
  \colhead{} & 
  \colhead{Clump A} & 
  \colhead{Clump B} &
  \colhead{Clump C}&
   \colhead{Clump D} & 
  \colhead{Clump E} \\
 }
  \startdata
$\Gamma(UVW2,F336W)$ & 1.33$\pm$0.201&1.35$\pm$0.478&1.81$\pm$0.268&1.64$\pm$0.167&1.28$\pm$0.218\\
$\Gamma(UVM2,F336W)$ & 1.57$\pm$0.186&2.07$\pm$0.286&2.47$\pm$0.293&2.39$\pm$0.225&1.81$\pm$0.497\\
$\Gamma(UVW1,F336W)$ & 1.26$\pm$0.107&1.51$\pm$0.266&1.44$\pm$0.161&1.36$\pm$0.215&1.31$\pm$0.178\\
\enddata
\label{t:uv_rela}
\end{deluxetable}

\begin{deluxetable}{ccc}
  \tabletypesize{\footnotesize}
 \tablecolumns{3}
  \tablecaption{Summary of Uncertainties in $\Gamma(n)$}
  \tablewidth{0pt}
  \tablehead{
  \colhead{Source} & 
  \colhead{Type} & 
  \colhead{Error} \\
  }
  \startdata
Foreground Galactic extinction & statistic & 10\% \\
Fraction of obscured star light & systematic & 5\% \\
Scattered light & systematic & 5\% \\
Unresolved Reddening Structure & systematic & 2\%\\
\enddata
\label{t:error}
\end{deluxetable}
\begin{deluxetable}{cccc}
  \tabletypesize{\small}
 \tablecolumns{4}
  \tablecaption{The Range of $A_{n}$/$A_{F547M}$ for Stellar Populations
    with Different Ages and Foreground Extinction Types}
  \tablewidth{0pt}
  \tablehead{
  \colhead{Filter} & 
  \colhead{Milky Way} & 
  \colhead{LMC} &
  \colhead{SMC} \\ 
  }
  \startdata
UVW2&1.60-2.65&1.56-2.56&1.68-3.29\\
UVM2&2.33-2.79&2.22-2.63&2.43-2.93\\
UVW1&1.67-2.20&1.63-2.12&1.78-2.38\\
F275W&1.92-2.03&1.87-1.96&2.09-2.21\\
F336W&1.62-1.62&1.61-1.61&1.73-1.74\\
F390M&1.47-1.48&1.42-1.42&1.54-1.55\\
F435W&1.30-1.33&1.26-1.29&1.34-1.37\\
F475W&1.17-1.20&1.15-1.18&1.19-1.23\\
F665N&0.79-0.80&0.86-0.86&0.76-0.76\\
F814W&0.58-0.59&0.71-0.71&0.56-0.57\\
F110W&0.33-0.33&0.37-0.38&0.24-0.25\\
F160W&0.20-0.20&0.20-0.20&0.15-0.16\\
\enddata
\label{t:alambda_evo}
\tablecomments{The metallicity of the background instantaneous population is
solar metallicity and the ages are from 100 Myr to 10 Gyr. }
\end{deluxetable}


\begin{thebibliography}{}
\bibitem[Beaton et al.(2007)]{bea07} Beaton, R.~L., Majewski, 
S.~R., Guhathakurta, P., et al.\ 2007, \apjl, 658, L91
\bibitem[Baes 
\& Gentile(2011)]{bae11} Baes, M., \& Gentile, G.\ 2011, \aap, 525, A136 
\bibitem[Bender et al.(2005)]{ben05} Bender, R., Kormendy, 
J., Bower, G., et al.\ 2005, \apj, 631, 280 
\bibitem[Bianchi et al.(1996)]{bia96} Bianchi, L., Clayton, 
G.~C., Bohlin, R.~C., Hutchings, J.~B., \& Massey, P.\ 1996, \apj,
471, 203 
\bibitem[Breeveld et al.(2010)]{bre10} Breeveld, A.~A., 
Curran, P.~A., Hoversten, E.~A.,  Koch, S., Landsman, W., Marshall,
F. E., Page, M. J., Poole, T. S., et al.\ 2010, \mnras, 406, 1687
\bibitem[Brown et al.(1998)]{bro98} Brown, T.~M., Ferguson, 
H.~C., Stanford, S.~A., \& Deharveng, J.-M.\ 1998, \apj, 504, 113 
\bibitem[Calzetti et al.(2000)]{cal00} Calzetti, D., Armus, 
L., Bohlin, R.~C., et al.\ 2000, \apj, 533, 682  
\bibitem[Calzetti(2001)]{cal01} Calzetti, D.\ 2001, \pasp, 
113, 1449 
\bibitem[Cardelli et al.(1989)]{car89} Cardelli, J.~A., 
Clayton, G.~C., \& Mathis, J.~S.\ 1989, \apj, 345, 245
\bibitem[Cartledge et al.(2005)]{car05} Cartledge, S.~I.~B., 
Clayton, G.~C., Gordon, K.~D., et al.\ 2005, \apj, 630, 355 
\bibitem[Crane et al.(1992)]{cra92} Crane, P.~C., Dickel, 
J.~R., \& Cowan, J.~J.\ 1992, \apjl, 390, L9
\bibitem[Dalcanton et al.(2012)]{dal12} Dalcanton, J.~J., 
Williams, B.~F., Lang, D., et al.\ 2012, \apjs, 200, 18 
\bibitem[Dong et al.(2011)]{don11} Dong, H., Wang, Q.~D., 
Cotera, A., et al.\ 2011, \mnras, 417, 114
\bibitem[Dong et al.(2014)]{don13} Dong, H. et al. (2014) in
  preparation
\bibitem[Draine(2003)]{dra03} Draine, B.~T.\ 2003, \apj, 598, 
1017 
\bibitem[Dressler 
\& Richstone(1988)]{dre88} Dressler, A., \& Richstone, D.~O.\ 1988, \apj, 324, 701
\bibitem[Elmegreen(1980)]{elm80} Elmegreen, D.~M.\ 1980, 
\apjs, 43, 37 
\bibitem[Fitzpatrick 
\& Massa(1990)]{fit90} Fitzpatrick, E.~L., \& Massa, D.\ 1990, \apjs, 72, 163 
\bibitem[Fitzpatrick(1999)]{fit99} Fitzpatrick, E.~L.\ 1999, 
\pasp, 111, 63 
\bibitem[Fitzpatrick(2004)]{fit04} Fitzpatrick, E.~L.\ 2004, 
Astrophysics of Dust, 309, 33
\bibitem[Fritz et al.(2011)]{fri11} Fritz, T.~K., Gillessen, 
S., Dodds-Eden, K., et al.\ 2011, \apj, 737, 73 
\bibitem[Garcia et al.(2010)]{gar10} Garcia, M.~R., Hextall, 
R., Baganoff, F.~K., et al.\ 2010, \apj, 710, 755 
\bibitem[Groves et al.(2012)]{gro12} Groves, B., Krause, O., 
Sandstrom, K., et al.\ 2012, \mnras, 426, 892 
\bibitem[Gordon 
\& Clayton(1998)]{gor98} Gordon, K.~D., \& Clayton, G.~C.\ 1998, \apj, 500, 816 
\bibitem[Gordon et al.(2003)]{gor03} Gordon, K.~D., Clayton, 
G.~C., Misselt, K.~A., Landolt, A.~U., \& Wolff, M.~J.\ 2003, \apj,
594, 279 
\bibitem[Gordon et al.(2009)]{gor09} Gordon, K.~D., 
Cartledge, S., \& Clayton, G.~C.\ 2009, \apj, 705, 1320 
\bibitem[Jacoby et al.(1985)]{jac85} Jacoby, G.~H., Ford, H., 
\& Ciardullo, R.\ 1985, \apj, 290, 136
\bibitem[Jedrzejewski(1987)]{jed87} Jedrzejewski, R.~I.\ 
1987, \mnras, 226, 747 
\bibitem[Jiang et al.(2013)]{jia13J} Jiang, P., Zhou, H., Ji, 
T., et al.\ 2013, \aj, 145, 157 
\bibitem[Jones(2004)]{jon04} Jones, A.~P.\ 2004, Astrophysics 
of Dust, 309, 347 
\bibitem[Kainulainen et 
al.(2009)]{kai09} Kainulainen, J.~T., Alves, J.~F., Beletsky, Y., et al.\ 2009, \aap, 502, L5 
\bibitem[Kormendy(1988)]{kor88} Kormendy, J.\ 1988, \apj, 
325, 128
\bibitem[Kormendy 
\& Bender(1999)]{kor99} Kormendy, J., \& Bender, R.\ 1999, \apj, 522, 772 
\bibitem[Lauer(1986)]{lau86} Lauer, T.~R.\ 1986, \apj, 311, 
34 
\bibitem[Lauer et al.(2012)]{lau12} Lauer, T.~R., Bender, R., 
Kormendy, J., Rosenfield, P., \& Green, R.~F.\ 2012, \apj, 745, 121
\bibitem[Li, Wang \& Wakker(2009)]{li09} Li, Z., Wang, Q.~D., 
\& Wakker, B.~P.\ 2009, \mnras, 397, 148 
\bibitem[Li et al.(2011)]{li11} Li, Z., Garcia, M.~R., 
Forman, W.~R., et al.\ 2011, \apjl, 728, L10 
\bibitem[Li et al.(2014)]{li13} Li, Z., et al., 2014, in preparation
\bibitem[Mathis 
\& Cardelli(1992)]{mat92} Mathis, J.~S., \& Cardelli, J.~A.\ 1992,
\apj, 398, 610 
\bibitem[Markwardt(2009)]{mar09} Markwardt, C. B., 2009, ASPC, 411, 251M
\bibitem[McConnachie et al.(2005)]{mcc05} McConnachie, A.~W., 
Irwin, M.~J., Ferguson, A.~M.~N., et al.\ 2005, \mnras, 356, 979 
\bibitem[Melchior et al.(2000)]{mel00} Melchior, A.-L., 
Viallefond, F., Gu{\'e}lin, M., \& Neininger, N.\ 2000, \mnras, 312,
L29  
\bibitem[Melchior 
\& Combes(2011)]{mel11} Melchior, A.-L., \& Combes, F.\ 2011, \aap,
536, A52 
\bibitem[Melchior 
\& Combes(2013)]{mel13} Melchior, A.-L., \& Combes, F.\ 2013, \aap, 549, A27 
\bibitem[Menten et 
al.(2007)]{men07} Menten, K.~M., Reid, M.~J., Forbrich, J., \& Brunthaler, A.\ 2007, \aap, 474, 515 
\bibitem[Massa et al.(1983)]{mas83} Massa, D., Savage, B.~D., 
\& Fitzpatrick, E.~L.\ 1983, \apj, 266, 662 
\bibitem[Misselt et al.(1999)]{mis99} Misselt, K.~A., 
Clayton, G.~C., \& Gordon, K.~D.\ 1999, \apj, 515, 128 
\bibitem[Morrissey et al.(2007)]{mor07} Morrissey, P., 
Conrow, T., Barlow, T.~A., et al.\ 2007, \apjs, 173, 682 
\bibitem[Nataf et al.(2013)]{nat13} Nataf, D.~M., Gould, A., 
Fouqu{\'e}, P., et al.\ 2013, \apj, 769, 88
\bibitem[Olsen et al.(2006)]{ols06} Olsen, K.~A.~G., Blum, 
R.~D., Stephens, A.~W., et al.\ 2006, \aj, 132, 271 
\bibitem[Peng(2002)]{pen02} Peng, C.~Y.\ 2002, \aj, 124, 294 
\bibitem[Phillips et al.(2013)]{phi13} Phillips, M.~M., 
Simon, J.~D., Morrell, N., et al.\ 2013, \apj, 779, 38 
\bibitem[Revnivtsev et al.(2010)]{rev10} Revnivtsev, M., van den Berg, M., Burenin, R., et al.\ 2010, \aap, 515, A49 
\bibitem[Roming et al.(2005)]{rom05} Roming, P.~W.~A., 
Kennedy, T.~E., Mason, K.~O., Nousek, J. A., Ahr, L., Bingham, R. E.,
Broos, P. S., Carter, M., M., et al.\ 2005, \ssr, 120, 95 
\bibitem[Rosenfield et al.(2012)]{ros12} Rosenfield, P., 
Johnson, L.~C., Girardi, L., et al.\ 2012, \apj, 755, 131
\bibitem[Rosolowsky(2007)]{ros07} Rosolowsky, E.\ 2007, \apj, 
654, 240
\bibitem[Rubin 
\& Ford(1971)]{rub71} Rubin, V.~C., \& Ford, W.~K., Jr.\ 1971, \apj, 170, 25 
\bibitem[Saglia et 
al.(2010)]{sag10} Saglia, R.~P., Fabricius, M., Bender, R., et al.\ 2010, \aap, 509, A61 
\bibitem[Schlafly 
\& Finkbeiner(2011)]{sch11} Schlafly, E.~F., \& Finkbeiner, D.~P.\ 2011, \apj, 737, 103 
\bibitem[Sumi(2004)]{sum04} Sumi, T.\ 2004, \mnras, 349, 193
\bibitem[Stark(1977)]{sta77} Stark, A.~A.\ 1977, \apj, 213, 
368 
\bibitem[Udalski(2003)]{uda03} Udalski, A.\ 2003, \apj, 590, 
284 
\bibitem[Valencic et al.(2004)]{val04} Valencic, L.~A., 
Clayton, G.~C., \& Gordon, K.~D.\ 2004, \apj, 616, 912 
\bibitem[V{\'a}zquez 
\& Leitherer(2005)]{vaz05} V{\'a}zquez, G.~A., \& Leitherer, C.\ 2005,
\apj, 621, 695
\bibitem[Walterbos 
\& Kennicutt(1988)]{wal88} Walterbos, R.~A.~M., \& Kennicutt, R.~C., Jr.\ 1988, \aap, 198, 61 
\end{thebibliography}
\end{document}